\input epsf.sty
\overfullrule=0pt

\def\ctr#1{\noindent{\hfill#1\hfill}}

\def\simless{\mathbin{\lower 3pt\hbox
   {$\rlap{\raise 5pt\hbox{$\char'074$}}\mathchar"7218$}}}   
\def\simgreat{\mathbin{\lower 3pt\hbox
   {$\rlap{\raise 5pt\hbox{$\char'076$}}\mathchar"7218$}}}   

\font\bigbf=cmbx10 scaled\magstep1

%

\ctr{\bigbf LIGHT PROPAGATION IN INHOMOGENEOUS UNIVERSES}

\medskip

\ctr{\bigbf III. DISTRIBUTIONS OF IMAGE SEPARATIONS}

\bigskip

\ctr{HUGO MARTEL$^1$, PREMANA PREMADI$^2$, AND RICHARD MATZNER$^{3,4}$}

\bigskip\bigskip

\noindent $^1$ Department of Astronomy, University of Texas, Austin, TX 78712

\noindent $^2$ Department of Astronomy \& Bosscha Observatory,
                 Bandung Institute of Technology, Indonesia

\noindent $^3$ Center for Relativity, University of Texas, Austin, TX 78712

\noindent $^4$ Department of Physics, University of Texas, Austin, TX 78712

\bigskip\bigskip\bigskip

\ctr{\bf ABSTRACT}

\medskip

Using an analytical model, we compute the distribution of image separations
resulting from gravitational lensing of distant sources, for 7
{\sl COBE\/}-normalized CDM models with various combinations of $\Omega_0$ and
$\lambda_0$. Our model assumes that multiple imaging
results from strong lensing by individual galaxies. We model galaxies
as nonsingular isothermal spheres whose parameters
are functions of the luminosity and morphological type, and
take into account the finite angular size of the sources. 
Our model neglects the contribution of the background matter
distribution to lensing, and assumes that lensing is entirely
caused by galaxies. To test the validity of this assumption,
we performed a series of ray-tracing
experiments to study the effect of the background matter on the
distribution of image separations. Our results are the following:
(1) The presence of the background matter tends to increase the image
separations produced by lensing galaxies, making the distributions of
image separations wider. However, this effect is rather small,
and independent of the cosmological model. (2)
Simulations with galaxies and background matter often produce a
secondary peak in the distribution of image separations at large
separations. This peak does not appear when the background
matter is excluded from the simulations. (3) The effect of the background
matter on the magnification distribution is negligible in low
density universes ($\Omega_0=0.2$) with small 
density contrast ($\sigma_8=0.4$),
but becomes very important as $\Omega_0$ and $\sigma_8$ increase, resulting
in a significant widening of the distribution. (4) Multiple imaging is
caused primarily by early-type galaxies (elliptical and S0's), with
a negligible contribution from spiral galaxies. (5) Our analytical model,
which has only 2 free parameters, is in good agreement with the
results of ray-tracing experiments, successfully reproducing the
distributions of image separations, and also the multiple-imaging
probability, for all cosmological models considered. (6) The analytical
model predicts that the distributions of image separations are
virtually indistinguishable for flat, cosmological constant models with
different values of $\Omega_0$. (7) For 
models with no cosmological constant, the distributions of
image separations do depend upon $\Omega_0$, but this dependence
is weak. We conclude that while the number of multiple-imaged sources
can put strong constraints on the cosmological parameters, the distribution
of image separations does not constrain the cosmological models in any
significant way, and mostly provides constraints on the structure of
the galaxies responsible for lensing.

\vfill

\ctr{Accepted for publication in {\sl The Astrophysical Journal}}

\bigskip\bigskip

\eject

%

\ctr{\bf 1.\quad INTRODUCTION}

\medskip

The multiple imaging of distant source is the most spectacular
manifestation of gravitational lensing. Since the discovery
of the original binary quasar 0957+561 (Walsh, Carswell, \& Weymann 1979), 
the number of multiply-imaged
sources have been steadily growing, and at present 64
sources with multiple images have been identified
(Kochanek et al. 1998)\footnote{$^5$}{See http://cfa-www.harvard.edu/castles}. 
Multiple-image 
systems are characterized by their total magnification, image brightness
ratio, and image separation. The distribution of these quantities, the
image separation in particular, can be used to study the properties of
the lenses, of the underlying cosmological background, and, to some
extent, of the sources. A common approach for studying gravitational
lensing consists of performing ray-tracing experiments, using a
multiple lens-plane algorithm (Schneider, Ehlers, \& Falco 1992,
hereafter SEF). These experiments predict numerous properties of
gravitational lenses that can be compared with observations to
constrain the cosmological models. We have designed our own version of
the multiple lens-plane algorithm (Premadi, Martel, \& Matzner 1998,
hereafter Paper~I). This
version includes the contributions to lensing caused by the distribution 
of background dark matter, as well as individual galaxies. We have used
this algorithm to perform the largest cosmological parameter survey
ever done in this field (Premadi et al. 2001, hereafter Paper~II).

In Paper~II, we introduced an analytical model for the distribution of
image separations. This model successfully reproduced the least noisy of
our numerically-generated distributions, the ones for which the number
of cases was large enough to be statistically significant. The goal of this
paper is to explore this analytical model in more detail,
to investigate its properties, its accuracy, and the
validity of the assumptions on which it is based.

Modeling the distribution of image separations analytically is not
a new idea. In their seminal paper, Turner, Ostriker, \& Gott
(1984, hereafter TOG) studied the image separations resulting from
galaxies modeled as point masses or singular isothermal spheres,
in universes with density parameter
$\Omega_0=0$ and $\Omega_0=1$. More recent studies include
Hinshaw \& Krauss (1987), Narayan \& White (1988), Paczy\'nski \& Wambsganss
(1989), Kochanek (1995), Porciani \& Madau (2000), 
Keeton, Christlein, \& Zabludoff (2000), Li \& Ostriker (2001),
Keeton (2001), and Takahashi \& Chiba (2001).
The main difference between these previous studies and our own is
in the choice of
galactic model. With the exception of Hinshaw \& Krauss (1987), these 
studies only considered galactic models with singular density profiles,
either point masses, singular isothermal spheres, or
NFW halos (see Navarro, Frenk, \& White 1996, 1997).
We model galaxies as nonsingular isothermal spheres with a finite core.
This is an important assumption that needs to be justified. Recent
large N-body simulations of structure formation in CDM models predict that
galaxies and clusters have a singular, or ``cuspy,'' density profile
which approaches a power law $\rho\propto r^{-n}$ at the center,
with the exponent $n$ ranging from 1 to 1.5 (Cole \& Lacey 1996;
Navarro et al. 1996, 1997; Tormen, Bouchet, \& White 1997;
Fukushige \& Makino 1997, 2001a, b; Moore et al. 1998, 1999; 
Huss, Jain, \& Steinmetz 1999; Ghigna et al. 2000; Jing \& Suto 2000;
Klypin et al. 2000. See, however, Kravtsov et al. 1998). 
However, these results are in conflict with several observations.
Dark-matter-dominated dwarf galaxies and low surface 
brightness galaxies exhibit rotation curves which imply mass profiles which 
are inconsistent with a singular shape, but are well fit by a mass profile
with a central core (see reviews by Primack et
al. 1998; Burkert \& Silk 1999; and Shapiro, Iliev, \& Raga 1999).
At the other extreme,
on the cluster scale, observations of strong gravitational lensing
of background galaxies by foreground clusters indicate the presence of a
finite-density core in the center of clusters (Tyson, Kochanski, \&
Dell'Antonio 1998). At the intermediate scale, HST observations of
massive early-type galaxies (Lauer et al. 1995) reveal that these
galaxies have brightness profiles that break from steep outer
power laws to shallower inner cusps. The inner profiles, inside the
break, remain singular, but the power-law exponents have a wide, bimodal
distribution, ranging from $n=0$ to $n=2.5$
(Gebhardt et al. 1996), in conflict with numerical simulations.

This conflict between observations and
simulations is a major challenge to the current CDM model of structure
formation, and some authors have even claimed that the CDM model is
ruled out. The most widely accepted viewpoint is that the observations are
correct, and that the N-body simulations are in disagreement with observations
because they are ignoring a crucial physical process. The two leading
candidates for this physical process are gasdynamics 
(see, e.g. El-Zant, Shlosman, \& Hoffman 2001) and self-interacting
dark matter (Spergel \& Steinhardt 2000), 
which both have the potential to eliminate the central
cusp in the density profile and reconcile simulations with observations.

In this paper, we take the viewpoint that in
the presence of such conflict, we should regard as ``realistic'' a galaxy
model that is in good agreement with the observations, even if it might be
in conflict with the simulations. Hence, we use a galaxy model (non-singular
isothermal sphere) which has a finite-density core. This is what
distinguishes this work from other analytical studies.
Of all these previous studies, the one that most closely resembles our work
is the study of Hinshaw \& Krauss (1987), the only one that
considers the effect of a finite-density core on the distribution
of image separations. These authors model galaxies
as nonsingular isothermal spheres which follow
a Schechter luminosity function. We use the same assumption in this paper.
Still, there are 5 important differences between this study and the one
of Hinshaw \& Krauss: (1) we consider different cosmological models, including
in particular flat models with a nonzero cosmological constant, (2) our
analytical
model corrects the lensing cross section to account for the finite size of the
source, (3) we include an observational
selection effect, by discarding multiple images if 
the image separation is too small for the images
to be resolved individually, either because the images overlap or
the angular spacing between them is below the 
resolution limit of the observations, (4)
while the study of Hinshaw \& Krauss was purely analytical, we compare
our analytical predictions with results of ray-tracing
experiments, (5) we investigate the effect of the background matter
on lensing by galaxies. 

The remainder of this paper is organized as follows. In \S2, we describe
the cosmological models considered in this paper. In \S3, we investigate the
effect of the background matter on the distributions of image separations
and magnifications. In \S4, we describe our analytical model for the 
distributions of image separations. Results are presented in \S5 and 
summarized in \S6.

\bigskip\smallskip

\ctr{\bf 2.\quad THE COSMOLOGICAL MODELS}

\medskip

In Paper~II, we considered 43 different {\sl COBE}-normalized
Tilted Cold Dark Matter models.
Each model was characterized by the value of the density parameter $\Omega_0$,
cosmological constant $\lambda_0$, Hubble constant $H_0$, and rms
density fluctuation $\sigma_8$. The tilt $n$ of the power spectrum
was adjusted in order to reproduce the desired value of $\sigma_8$ for
any particular combination of $\Omega_0$, $\lambda_0$, and $H_0$. We are 
considering the same models in this paper, and use, for our 
analytical calculations, the same galaxy distributions that were generated 
for the calculations presented in Paper~II. As we will see in \S3, our
analytical model depends only
on the intrinsic properties of galaxies, and not their level of clustering.
Hence, the resulting separation distributions are independent of $\sigma_8$,
and galaxy distributions taken from
models with different values of $\sigma_8$ can be combined. Furthermore,
the analytical model turns out to be independent of $H_0$.
This reduces
the 43 four-parameter models of Paper~II to 7 two-parameter models. The
values of the parameters $\Omega_0$ and $\lambda_0$ for these 7 models
are listed in the second and third columns of Table 1.

\bigskip\smallskip

\ctr{Table 1: The Cosmological Models}

\ctr{\vbox{\halign{
\strut#\hfil&\quad\hfil#\hfil&\quad\hfil#\hfil&\quad\hfil#\hfil&
\quad\hfil#\hfil&\quad\hfil#\hfil&\quad\hfil#\hfil\cr
\noalign{\bigskip\hrule\smallskip\hrule\medskip}
Model & $\Omega_0$ & $\lambda_0$ & $\sigma_8$ & 
\% Ellipticals$^{\rm a}$ & \% S0's$^{\rm a}$ & \% Spirals$^{\rm a}$ \cr
\noalign{\smallskip\hrule\smallskip}
O1 & 0.2 & 0.0 & $0.3-0.7$ & $11-13$ & $27-34$ & $53-62$ \cr
L1 & 0.2 & 0.8 & $0.6-1.0$ & $12-14$ & $32-35$ & $51-56$ \cr
O2 & 0.5 & 0.0 & $0.8-1.0$ & $13-14$ & $34-36$ & $50-53$ \cr
L2 & 0.5 & 0.5 & $0.8-1.0$ & $13-14$ & $34-36$ & $51-53$ \cr
O3 & 0.7 & 0.0 & $0.9-1.1$ & $13-14$ & $35-36$ & $50-52$ \cr
L3 & 0.7 & 0.3 & $0.9-1.1$ & $13-14$ & $35-36$ & $50-52$ \cr
E  & 1.0 & 0.0 & $0.9-1.3$ & $13-14$ & $35-36$ & $49-51$ \cr
\noalign{\smallskip\hrule\smallskip}
\multispan7 $^{\rm a}$ Percentages are rounded to the nearest integer. \hfil\cr
}}}

\bigskip

Strictly speaking, this reduction from 43 models to 7 models
is not quite correct. The galaxy distributions actually
depend upon $\sigma_8$, but that dependence is indirect and quite weak. 
In the original algorithm of Jaroszy\'nski (1991,1992), the morphological
types of galaxies were chosen randomly. In Papers~I and~II, we improved 
the algorithm by choosing the morphological types according to the
observed morphology-density relation (Dressler 1980; Postman \& Geller 1984),
locating more early-type galaxies and fewer spiral galaxies in dense regions.
Since the density contrast of dense regions depends directly upon $\sigma_8$,
the relative number of early-type and spiral galaxies varies with 
$\sigma_8$, and this should affect the galaxy distributions.
The last 4 columns of Table~1
list the range of values of $\sigma_8$, and of the percentages of ellipticals, 
S0's, and spiral galaxies, respectively, for the various cosmological
models. Varying the
value of $\sigma_8$ does not affect dramatically the proportions of galaxies
of various types. The largest fluctuations are for the model $\Omega_0=0.2$,
$\lambda_0=0$, $H_0=75$, for which $\sigma_8$ varies from 0.3 to 0.7,
and the corresponding fraction of spiral galaxies varies from 56\% to
62\%. We have compared distributions predicted by models differing only
by the value of $\sigma_8$, and found them to be nearly undistinguishable.
Therefore, we are justified to combine models with different values of 
$\sigma_8$.

\bigskip\smallskip

\ctr{\bf 3.\quad THE EFFECT OF THE BACKGROUND MATTER}

\medskip

A crucial assumption of our analytical model is that gravitational
lensing is essentially caused by individual galaxies, and that the effect of
the background matter is unimportant and can be neglected. This is contrary to 
the widespread opinion that gravitational lensing by galaxies can be
significantly affected by the presence of the background matter.
This opinion is based primarily on the work of TOG.
These authors calculated the effect of the
background matter on lensing analytically, and provided formulae to
correct the image separation and magnification produced by a galaxy for
the presence of the background matter (TOG, eqs.~[2.36] and~[2.37]).
To obtain this analytical result, they had to make 3 simplifying assumptions:
(1) the galaxies are modeled as singular isothermal spheres, with a 
$\rho\propto r^{-2}$ density profile, (2) the background matter is 
approximated as a sheet of matter of constant surface density $\Sigma$ and
(3) the source is aligned with the lensing galaxy. While assumption (3)
is probably of little consequence for the image separations,\footnote{$^6$}{The 
image separation is
actually independent of the position of the source in the
absence of the background matter, for a singular isothermal sphere.}
the other assumptions are not.
A singular isothermal sphere has a density profile that is much steeper
than the profiles obtained by numerical simulations, which are
themselves steeper than the observed density profiles of galaxies and
clusters. Also, a uniform sheet of constant surface density is a very crude
representation of the actual structures that form out of Gaussian random
noise initial conditions in a CDM universe. TOG, in their Appendix~B, warn 
their readers that in the real universe, there would be overdense and
underdense regions (``sheets'' and ``holes'') along the line of sight,
and their effects would partly cancel out.

To estimate the effect of the background matter in the context of
a CDM universe,
we went back to the ray-tracing experiments presented in
Paper~II, which included the effects of both galaxies and background matter.
We then modified our multiple lens-plane algorithm to take out the contribution
from the background matter, and did 3 new series of experiments, for
3 different cosmological models: 
an open model ($\Omega_0=0.2$, $\lambda_0=0$, 
$H_0=75\,\rm km\,s^{-1}Mpc^{-1}$, $\sigma_8=0.4$)
a flat, cosmological constant model ($\Omega_0=0.2$, $\lambda_0=0.8$, 
$H_0=65\,\rm km\,s^{-1}Mpc^{-1}$, $\sigma_8=0.8$)
and an Einstein-de~Sitter model ($\Omega_0=1$, $\lambda_0=0$, 
$H_0=65\,\rm km\,s^{-1}Mpc^{-1}$, $\sigma_8=0.9$). In this section, we compare
the results of the experiments with and without the contribution
of the background matter included.

\bigskip\smallskip

\ctr{3.1.\quad The Distribution of Image Separations}

\medskip

Figure~1 shows the distributions of image separations for the 3 
cosmological models considered. The solid lines are the distributions
obtained when the effect of galaxies and background matter are
both included (these histograms are taken directly from
Paper~II). The dotted lines are the distributions obtained when only the
effect of galaxies is included. All three panels show a similar trend:
when only galaxies are included, the peaks of the distributions are higher,
and the distributions are narrower. The presence of the background matter
lowers the counts at separations $s\simless2''$ and increases them at
separations $s\simgreat2''$, producing a tail that extends to high separations.
Notice that the secondary peak in the distribution at $s=5''$ for the
flat model (middle panel) disappear when the background matter is removed.
We speculated in Paper II that this secondary peak, present for many
cosmological models, resulted from the combined effect of galaxies and
background matter. The results plotted in Figure~1 support this idea.

We computed the mean $\bar s$, standard deviation
$\sigma_s$, and skewness of the
distributions plotted in Figure~1. The results are listed in Table~2,
where an $\times$ in the second column indicates that the background matter
was included ($\bar s$ and $\sigma_s$ are in arc seconds; the skewness is
dimensionless). The values of $\bar s$ and $\sigma_s$ are very similar 
among the 3 cases with background matter, and also among the 3 cases
without. The presence of the background matter increases $\bar s$ by about
10\% and $\sigma_s$ by about 15\%. The skewness is definitely positive, and
tends to increase when the background matter is included, because of the 
presence of the high-separation tail. These results are consistent with
claims that the effect of the background matter on image separation
is of order 20\% or less (Bernstein \& Fischer 1999; Romanowsky \& Kochanek
1999). Keeton et al. (2000) showed that the dependence of the luminosity
function on environment, which we neglect in this paper, decreases the
proportion of high-mass galaxies relative to dwarfs in dense environments,
and that this effect nearly cancels the effect of the background matter,
making the distributions of image separations essentially independent of
environment.

\epsfxsize=12cm
\vskip-0.6cm
\hskip1.2cm\epsfbox{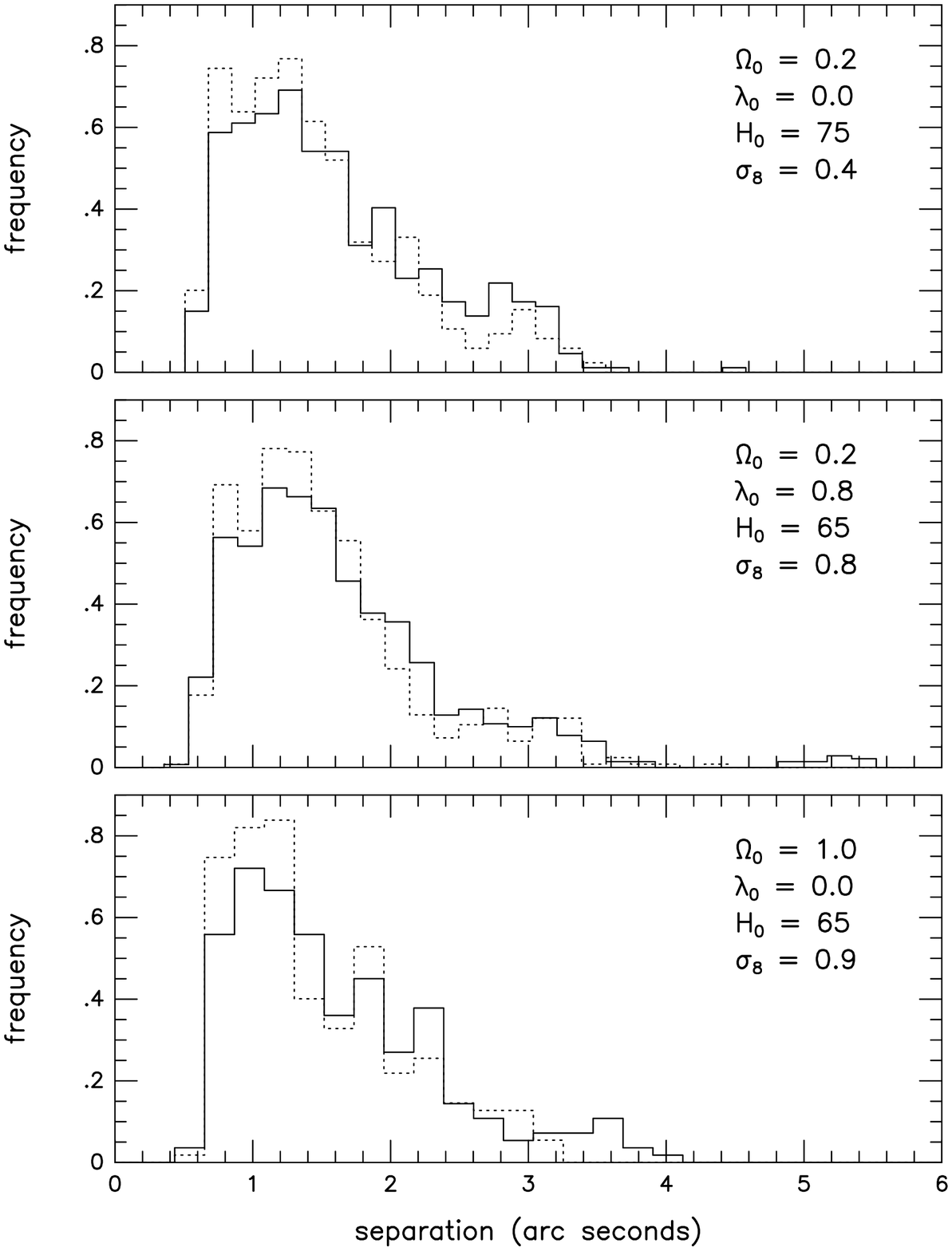}
\vskip-0.6cm

{\narrower\noindent
Fig. 1: Distribution of image separations in arc seconds for
3 different cosmological models.
The solid lines show the distribution obtained when 
the presence of the background matter is taken into account.
The dotted lines show the distribution obtained when 
the presence of the background matter is ignored. Both distributions
were computed using ray-tracing experiments.
\bigskip\smallskip}

Figure~1 and Table~2
yield two important results concerning the distributions of
image separations: (1) The effect of the background matter on the
distributions of image separations produced by galaxies is essentially
independent of the cosmological model. Not only do we observe the same trend
in all panels of Figure~1, but the effect of
the background matter is of the same order for all models.
(2) The effect of the background matter is far from spectacular.
The only significant effect that might be observable is the 
high-separation tail. In any case, these results show that the effect of the
background matter is small and it is not unreasonable to neglect it, 
as we do in our analytical model.

\bigskip\smallskip

\ctr{Table 2: Statistics of the Distributions of Image Separations}

\ctr{\vbox{\halign{
\strut\hfil#\hfil&\quad\hfil#\hfil&\quad\hfil#\hfil&
\quad\hfil#\hfil&\quad\hfil#\hfil\cr
\noalign{\bigskip\hrule\smallskip\hrule\medskip}
Model $(\Omega_0,\lambda_0)$ & Background & $\bar s$ & $\sigma_s$ & Skewness \cr
\noalign{\smallskip\hrule\smallskip}
(0.2,0.0) & $\times$ & 1.60 & 0.71 & $0.81\pm0.17$ \cr
(0.2,0.0) & $-$      & 1.47 & 0.64 & $0.99\pm0.17$ \cr
(0.2,0.8) & $\times$ & 1.64 & 0.80 & $1.67\pm0.14$ \cr
(0.2,0.8) & $-$      & 1.52 & 0.67 & $1.22\pm0.15$ \cr
(1.0,0.0) & $\times$ & 1.61 & 0.74 & $1.03\pm0.24$ \cr
(1.0,0.0) & $-$      & 1.46 & 0.62 & $0.83\pm0.24$ \cr
\noalign{\smallskip\hrule}
}}}

\bigskip

\ctr{3.2.\quad The Magnification Distribution}

\medskip

Using the calculations presented in this section, we estimate
the effect of the background matter on the distribution of magnifications,
even though this is not the main focus of this paper, which is
concerned only with the distribution of image separations. We include 
this subsection for completeness. Figure~2 shows the distributions
of magnifications for the 3 cosmological models considered. As in
Figure~1, the solid curves show the results when both galaxies and
background matter are included (they are also taken directly from
Paper~II), while the dotted curves show the distributions when only
galaxies are included. There is a major difference between these
results and the ones shown in Figure~1. While the effect of
the background matter on the distribution of image separations is
model-independent, the effect on the magnification
distribution strongly depends on the cosmological model. In the
top panel, for the open model, the two curves are nearly undistinguishable,
indicating that the effect of the background matter is totally negligible.
For the two other models, the presence of the background matter results
in a widening of the distribution, which is moderate for the flat
model (middle panel) but very important for the Einstein-de~Sitter model
(bottom panel).

Since the focus of this paper is on the distribution of image separations,
we will postpone a detailed study of the magnification distribution to
another paper. But let us speculate on the origin of the phenomenon
revealed by Figure~2. In the filled-beam approximation, which we use in
these calculations, the background matter would have absolutely no effect
on lensing if it was uniform. The density fluctuations in the background matter
are responsible for deflecting light, and the magnitude of these fluctuations
are measured by the parameter $\sigma_8$. Thus, we expect the effect of
the background matter to be larger for the flat and Einstein-de~Sitter
models, which have $\sigma_8=0.8$ and 0.9 respectively, than for the
open model, which has $\sigma_8=0.4$. This is indeed the case, but it does
not explain why the effect of the background matter for the open model
is totally negligible. We speculate that the explanation resides in the
structure formation process. In cosmological model such as CDM, structures
form hierarchically through mergers. As the universe evolves, clusters become
more massive, voids become deeper, and the actual number of voids and clusters
decreases. The value of $\sigma_8$ measures the level of clustering, and
is therefore related to the stage that this hierarchical formation process
has reached. In models with small $\sigma_8$, not only the density fluctuations
are small compared with model with higher $\sigma_8$, but in addition the
overdense and underdense regions have a smaller physical size (since the 
hierarchical merging process is not as well-advanced), and therefore there
should be a very large number of overdense and underdense regions along
any line of sight, leading to a near-perfect cancellation. A model with
larger $\sigma_8$ has fewer clusters and voids along the line of sight, and
a cancellation is less likely.
This idea certainly deserves more investigation.

Finally, let us point out that the difference between the flat model (middle
panel) and the Einstein-de~Sitter model (bottom panel) results primarily from
the different values of $\Omega_0$. The Einstein-de~Sitter model has nearly 5
times more background matter than the flat model.

\epsfxsize=12cm
\vskip-0.6cm
\hskip1.2cm\epsfbox{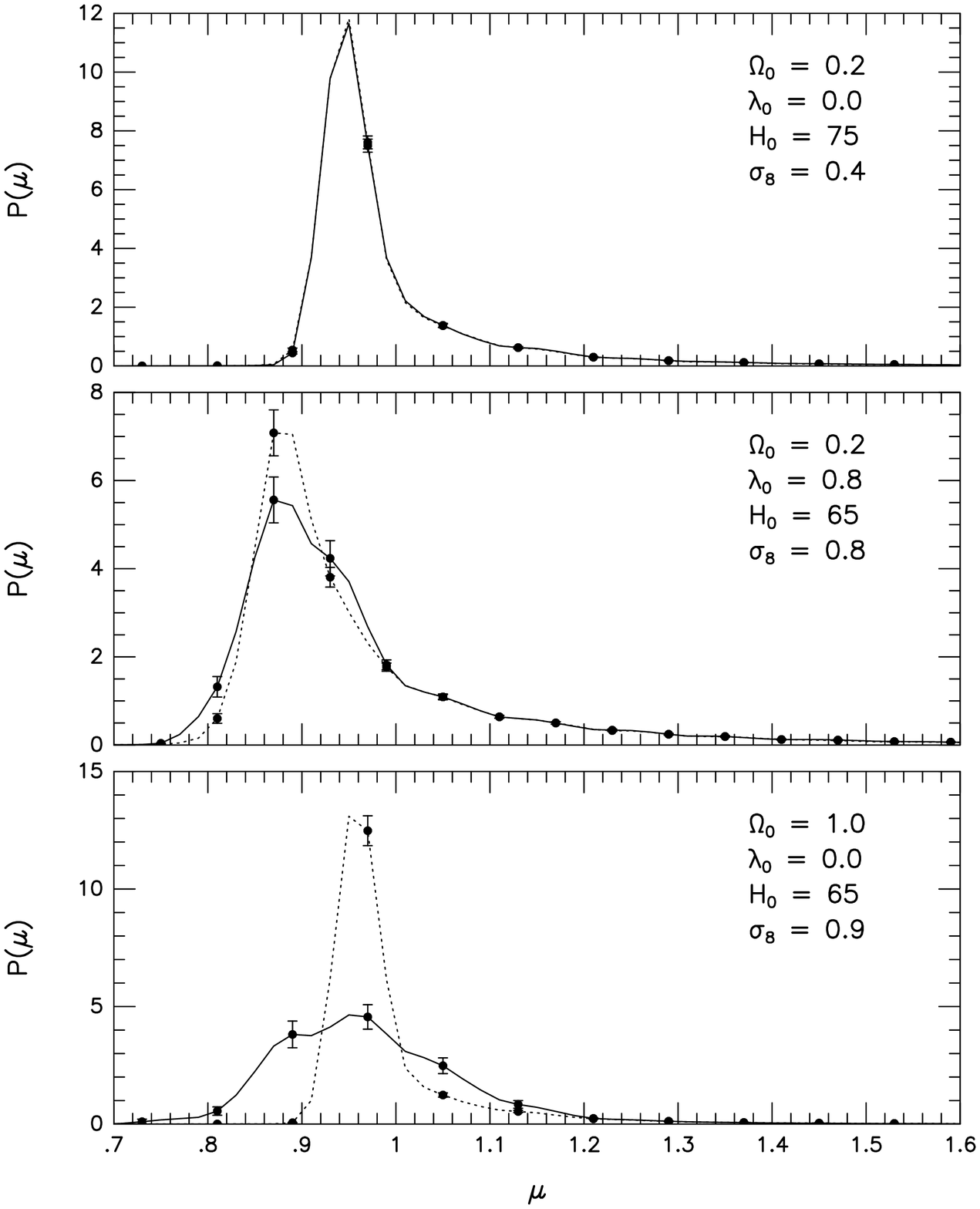}
\vskip-0.6cm

{\narrower\noindent
Fig. 2: Distribution of magnifications for
3 different cosmological models.
The solid curves show the distribution obtained when 
the presence of the background matter is taken into account.
The dotted curves show the distribution obtained when 
the presence of the background matter is ignored. 
Error bars indicate $1\sigma$ uncertainties. Both distributions
were computed using ray-tracing experiments.
\bigskip\smallskip}

\bigskip\smallskip

\ctr{\bf 4.\quad ANALYTICAL MODEL OF THE  DISTRIBUTION OF IMAGE SEPARATIONS}

\medskip

In this section, we describe the various assumptions on which our
analytical model is based. The validity of these assumptions is
discussed in \S6 below.

\bigskip\smallskip

\ctr{4.1.\quad The Galactic Models}

\medskip

We assume that the galaxy luminosities follow
a Schechter luminosity function,
$$n(L)dL={n_*\over L_*}\left({L\over L_*}\right)^{\alpha}e^{-L/L_*}dL\,,
\eqno{(1)}$$

\noindent
where $n(L)$ is the number density of galaxies per unit luminosity. 
We use the values
$\alpha=-1.10$, $n_*=0.0156\,h^{3}{\rm Mpc}^{-3}$, and $L_*=1.3\times
10^{10}h^{-2}L_{\odot}$, where $h$ is the
Hubble constant in units of $\rm 100\,km\,s^{-1}Mpc^{-1}$ 
(Efstathiou, Ellis, \& Peterson 1988). We also introduce a minimum
luminosity $L_{\min}=0.01L_*$ to prevent the total number of galaxies
from diverging. The corresponding present number density
and luminosity density are
$n_0=0.0808\,h^3\rm Mpc^{-3}$ and $j_0=2.13\times10^8hL_\odot
\,\rm Mpc^{-3}$, respectively.
We adopt the galaxy models described by Jaroszy\'nski (1991, 1992).
Each galaxy is modeled by a truncated, non-singular isothermal sphere,
whose parameters depend upon the galaxy luminosity and morphological type.
The projected surface density of each galaxy is given by
$$\sigma(r)={v^2\over4G(r^2+r_c^2)^{1/2}}\,,\eqno{(2)}$$

\noindent where $r$ is the projected distance from the center.
The parameters $r_c$ and $v$ are the core radius
and rotation velocity, respectively, and are given by
$$\eqalignno{
r_c&=r_{c0}\left({L\over L_*}\right)\,,&(3)\cr
v&=v_0\left({L\over L_*}\right)^\gamma\,,&(4)\cr}$$

\noindent where the parameters $r_{c0}$, $v_0$, and $\gamma$ are
given in Table~3.
We use a Monte-Carlo method to generate for each galaxy a luminosity
$L\geq L_{\min}$, with a probability $P(L)$ proportional to
$n(L)$. 

\bigskip\smallskip

\ctr{Table 3: Galaxy Parameters}

\ctr{\vbox{\halign{
\strut#\hfil&\quad\hfil#\hfil&\quad\hfil#\hfil&\quad\hfil#\hfil\cr
\noalign{\bigskip\hrule\smallskip\hrule\medskip}
Type & $r_{c0}$ ($h^{-1}\rm kpc$) & $v_0$ ($\rm km\,s^{-1}$) & $\gamma$ \cr
\noalign{\smallskip\hrule\smallskip}
Elliptical & 0.1 & 390 & 0.250 \cr
S0         & 0.1 & 357 & 0.250 \cr
Spiral     & 1.0 & 190 & 0.381 \cr
\noalign{\smallskip\hrule}
}}}

\bigskip\smallskip

\ctr{4.2\quad The Analytical Model}

\medskip

To compute analytically the distribution of image
separations, we make the following assumptions:

(1) Lensing is entirely caused by galaxies; we ignore the effect of
the background matter. Based on the results presented in \S3.1, this
is a reasonable assumption.

(2) Each galaxy acts as if it was alone;
we ignore the tidal effects of nearby galaxies, and the possibility of 
lensing events involving several galaxies. We can justify this
assumption as follows. Lensing is produced by the most
massive galaxies, which are also the most luminous. These galaxies are
quite rare. Equation~(1), with our assumed cutoff $L_{\rm min}=0.01L_*$ 
predicts that only 2\% of galaxies have a luminosity $L>L_*$. Hence, a galaxy
massive enough to produce multiple images is likely to be surrounded by
much less massive galaxies, whose tidal influence will be at most a small
perturbation.

(3) Galaxies are modeled as nonsingular isothermal spheres, 
as described in \S4.1.

With these assumptions, the
problem is reduced to studying lensing by isolated, nonsingular isothermal
spheres. The geometry of this problem is illustrated in Figure~3.
The optical axis goes through the observer and the center of the
lensing galaxy, $r_c$ is the core radius of the galaxy, $\eta$ is the distance
between the source and the optical axis, and the quantities $D_L$,
$D_{LS}$, and $D_S$ are the angular diameter distances
between observer and lens,
lens and source, and observer and source, respectively. All properties of 
this lensing system can be expressed as functions of dimensionless ratios
between $r_c$, $\eta$, $D_L$, $D_{LS}$, and $D_S$, and the
dimensionless ratio $v/c$, where $r_c$ and $v$ are given by
equations~(3) and~(4), respectively, and $c$ is the speed of light.

\epsfxsize=12cm
\hskip1.5cm\epsfbox{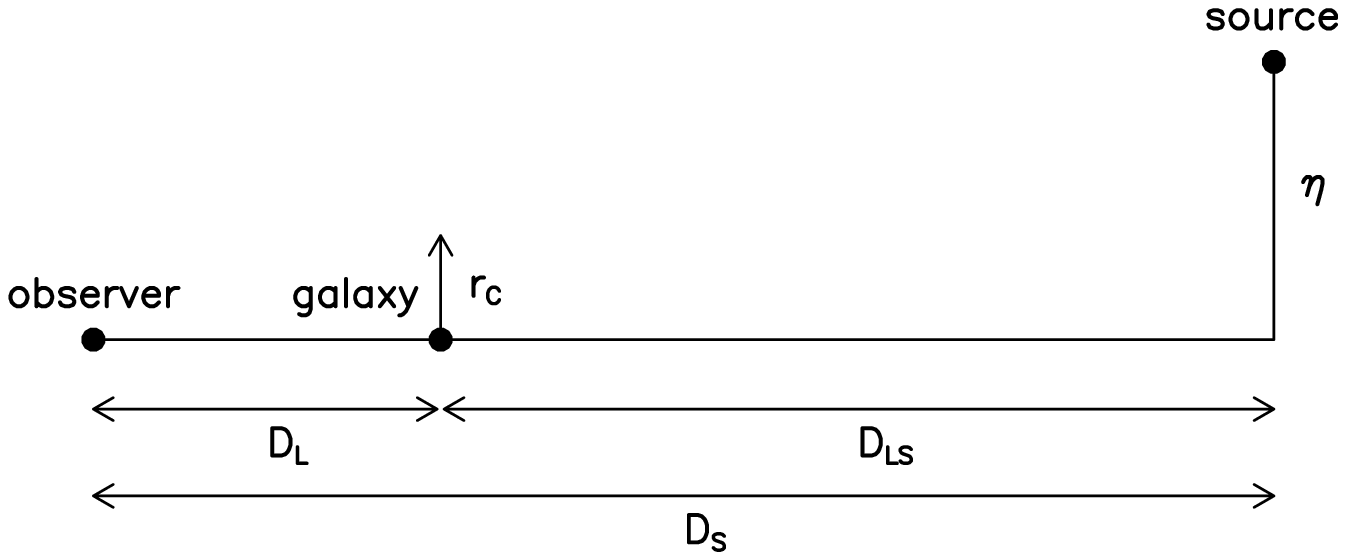}
\vskip-1cm

{\narrower\noindent
Fig. 3: The lensing geometry: the dots indicate the location of the 
observer, lensing galaxy, and source. $r_c$ is the core radius of the galaxy, 
and $\eta$ is the distance between the source and the optical axis.
The angular diameter distances $D_L$, $D_{LS}$, and $D_S$ are also indicated.
\bigskip\medskip}

(4) We neglect any spatial correlation between the sources and the lens.
The probability that a particular galaxy will produce multiple
images is then proportional to its angular cross section for multiple
imaging. However, we shall include a correction to this
cross section to account for the finite angular size of the sources
(see below).

(5) To build a realistic
distribution, we must impose limits on the smallest possible image
separation that allows individual images to be resolved. Throughout
this paper, we assume that
sources have an angular diameter $2\alpha_S=1''$, where
$\alpha_S$ is the angular radius of the source. The smallest possible
image separation is then of order $\alpha_S=0.5''$. To illustrate this, we
reproduce in Figure~4 a numerically-generated double
image, taken from Paper~II (Fig.~13c in that paper). 
The brightness ratio between the two images is very high, and the
total magnification is not much larger than unity. This double image
has about the same size as the source, and the image separation,
measured between image centers, is therefore of order
half the source diameter. However, the gap between the images is very
small and might be difficult to resolve observationally. To
take this selection effect into account, we introduce a free parameter $f>1$,
and we assume that at separations $s<\alpha_S$
the individual images can never be resolved, that at separations
$s>f\alpha_S$ they can always be resolved, and that at separations 
$\alpha_S<s<f\alpha_S$
they can sometimes be resolved, with a probability that varies
linearly from 0 to 1 between $s=\alpha_S$ and $s=f\alpha_S$. 
The probability of
``resolvability'' $P(s)$ is therefore given by
$$P(s)=\cases{
0\,, & $s<\alpha_S\,;$ \cr
(s-\alpha_S)/\alpha_S(f-1)\,, & $\alpha_S<s<f\alpha_S\,;$ \cr
1\,, & $s>f\alpha_S\;.$ \cr}\eqno{(5)}$$

\noindent Throughout this paper, we assume $\alpha_S=0.5''$ and $f=2$.

\epsfxsize=5cm
\hskip5cm\epsfbox{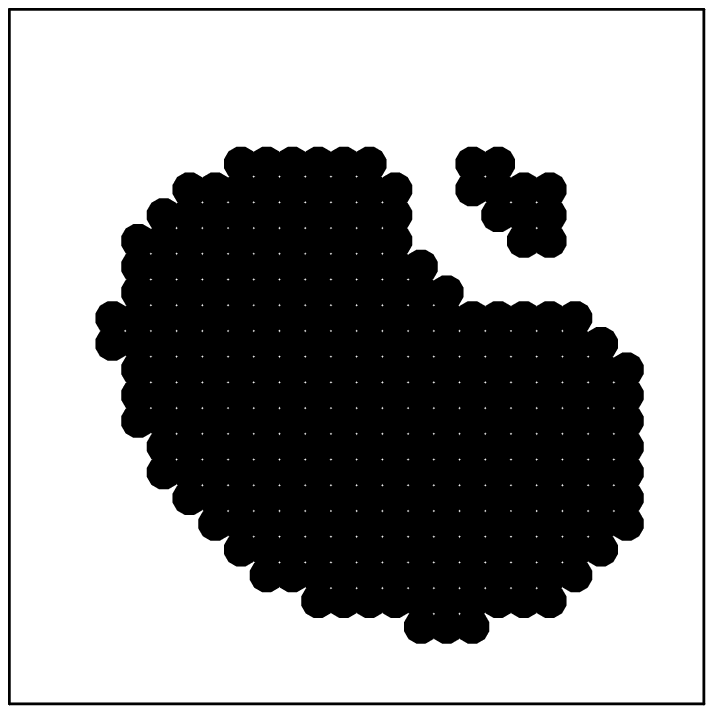}

{\narrower\noindent
Fig. 4: Double image with low magnification and high brightness ratio.
The image separation, measured between image centers, is of order the
source angular radius. This figure is reproduced from Paper~II.
\bigskip\medskip}

Gravitational lensing by isolated, nonsingular isothermal
spheres is an important problem, for which
several analytical results have been derived
(Dyer 1984; Hinshaw \& Krauss 1987; Blandford \& Kochanek 1987; 
Kochanek \& Blandford 1987). We can directly apply these results to
our analytical model.
For each galaxy, we introduce a length scale $\xi_0$, defined by
$$\xi_0=2\pi\left({v\over c}\right)^2{D_LD_{LS}\over D_S}\eqno{(6)}$$

\noindent (SEF, eq.~[8.34a], with $v=\sqrt{2}\sigma_v$). 
We use this parameter to 
rescale the core radius and the source position, as follows,
$$x_c=r_c/\xi_0\,,\qquad y=\eta/\xi_0\,.\eqno{(7)}$$

\epsfxsize=15cm
\vskip-6.5cm
\hskip0.3cm\epsfbox{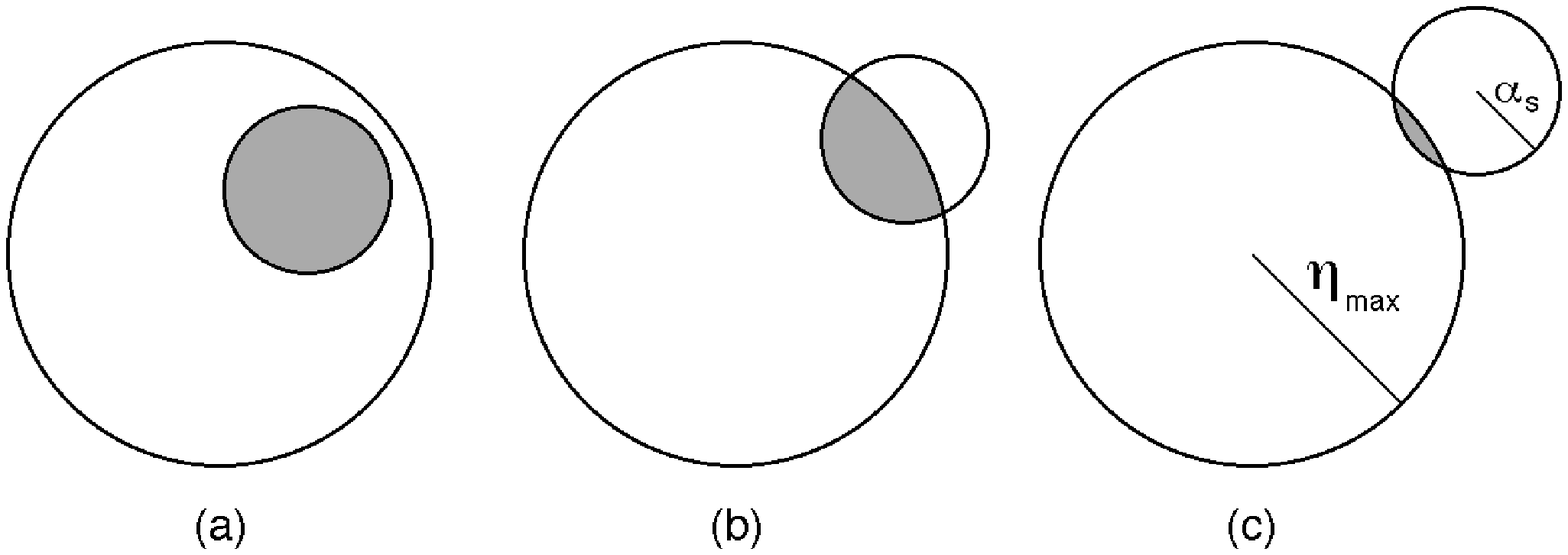}
\vskip-6.5cm

{\narrower\noindent
Fig. 5: Cross section for multiple imaging. The large and
small circles represent the cross section and the source,
respectively. The gray area indicates the part of the source that 
produces multiple images. $\eta_{\max}/D_S$ and $\alpha$ are
angular radii.
\bigskip\medskip}

\noindent
We will refer to $x_c$ as the {\it scaled core radius}.
We also define, for $x_c<1$, a critical radius 
$y_r=(1-x_c^{2/3})^{3/2}$. The nonsingular isothermal 
sphere has the following properties (SEF, \S12.2.3): (1) If $x_c\geq1$, the
source has only one image. (2) If $x_c<1$ the source has
one image if $y\geq y_r$, and 3 images if $y<y_r$. 
Hence, each lens which
satisfies the condition $x_c<1$ has an angular cross section for 
multiple imaging $\sigma_{\rm m.i.}$ given by
$$\sigma_{\rm m.i.}=\pi\left({\eta_{\rm max}\over D_S}\right)^2
=\pi\left({y_r\xi_0\over D_S}\right)^2\,.\eqno{(8)}$$

This expression is valid only for point sources. If the sources have a finite
size, as we assume in our model, part of the source might be located inside
the cross section even if the center of the source is not. 
This is illustrated in Figure~5, for 3 different locations of the source.
In each panel, the gray area indicates the fraction of the source that can
produce multiple images. We define the {\it effective}
cross section $\sigma_{\rm m.i.}^{\rm eff}$ as
$$\sigma_{\rm m.i.}^{\rm eff}
=\pi\left({\eta_{\rm max}\over D_S}+\zeta\alpha_S\right)^2
=\pi\left({y_r\xi_0\over D_S}+\zeta\alpha_S\right)^2\,,\eqno{(9)}$$

\noindent where
$\zeta$ is a tunable parameter. In principle, we should use $\zeta=1$, 
since overlap occurs whenever the angular separation between the
center of the source and the optical axis is less than
$\eta_{\max}/D_S+\alpha_S$.
However, if 
$\eta/D_S\simless\eta_{\max}/D_S+\alpha_S$, only a very small
fraction of the source overlaps with the cross section given by
equation~(8), as Figure~5c shows. 
Only that part of the source will form multiple
images, and these images will be very faint, possibly too faint to be 
resolved by observers, and also by ray-tracing experiments.\footnote{$^7$}{In 
Paper~II,
we discarded from the analysis any image composed of less than 5 light rays,
corresponding to a magnification less than 0.026.} We introduce the parameter
$\zeta$ to take this selection effect into account. We found that $\zeta=0.5$
is a good compromise, and leads to results that are in good agreement
with the results of Paper~II (see \S5.1 below).

If a galaxy modeled as a nonsingular isothermal
sphere produces multiple images (that is, 3 images), the angular separation
$s$ between the two outermost images depends upon the source position $\eta$,
and has a maximum value given by
$$s={2\xi_0\over D_L}(1-x_c^2)^{1/2}\,.\eqno{(10)}$$

\noindent Hinshaw \& Krauss (1987), and
Cheng \& Krauss (1999) showed that the dependence of $s$ on $\eta$
is weak. Following the suggestion made by SEF (p.~396), we will assume that
whenever multiple images occur, the image separation is
equal to this maximum value. 
We can then compute the distribution of image 
separations by including all galaxies which satisfy the condition $x_c<1$.
We give to each galaxy $j$ a weight $w_j$ equal to 
$$w_j={(\sigma_{\rm m.i.}^{\rm eff})_jP(s_j)\over C(z_j)}\,.\eqno{(11)}$$

\noindent where $s_j$ is the image separation produces by galaxy $j$ according
to equation~(10), $P(s_j)$ and $(\sigma_{\rm m.i.}^{\rm eff})_j$
are given by equations~(5) and ~(9), respectively, and
the redshift-dependent quantity $C(z)$ is the solid angle on
the sky at redshift $z$
that our sample of galaxies covers. This factor is given by
$C(z)=[L_{\rm box}/(1+z)/D_L(z)]^2$, where $L_{\rm box}=128\,\rm Mpc$ is the
comoving size of the computational volume in which the galaxy 
sample was generated (see Paper~II). The probability that a random source
will have resolvable multiple images is then given by adding up all the 
weights,
$$P_{\rm m.i.}=\sum_{\hbox{all $j$}}w_j\,.\eqno{(12)}$$

\bigskip\smallskip

\ctr{4.3.\quad Contributions of the Various Galaxy Types}

\medskip

Before computing the actual distributions of image separations, we
first estimate the relative contributions of high-mass and low-mass
galaxies, and of the various morphological types.
The usual assumption is that more massive galaxies produce larger
image separations, and for galaxies modeled as singular isothermal spheres,
this is certainly true. For galaxies with a finite core, this is only
true up to a point. In equation~(10), $\xi_0\propto v^2\propto L^{2\gamma}$,
and therefore in the limit $x_c\ll1$, the image separation increases with $L$.
However, combining equations~(3), (4), (6), and~(7), we get
$x_c\propto r_c/v^2\propto L^{1-2\gamma}$. Since $\gamma<1/2$ for all
morphological types (see Table~3), $x_c$ increases with $L$. Eventually, 
the factor $(1-x_c^2)^{1/2}$ in equation~(10) dominates, and the
image separation decreases, until $x_c=1$ and all images merge.
Hence, there are two characteristic luminosities in this problem:
the luminosity $L_s$ for which the image separation $s$ is maximum,
and the luminosity $L_1$, larger than $L_s$, for which $x_c=1$ and
only one image is produced (of course, $L_s$ and $L_1$ are functions of
the galaxy and source redshifts, the morphological type, and the
cosmological model through the dependences on $D_L$, $D_{LS}$, $D_S$,
and $\gamma$). To illustrate this, we plot in Figure~6 the image separation
produced by a lensing galaxy at redshift $z_L=0.3$, in a universe with
$\Omega_0=0.3$, $\lambda_0=0.7$, $H_0=70\rm\,km\,s^{-1}Mpc^{-1}$
(as usual, we assume a source redshift $z_S=3$). The top panel
shows the image separations produced by a spiral galaxy. The separation
increases with $L$ and reaches a maximum for $L=L_s$, then, the
effect of the finite core radius becomes important, and the separation
decreases until $L=L_1$ and the images merge together. The bottom panel
shows the image separation produced by a S0 galaxy. The core radius
is much smaller, and this pushes the values of $L_s$ and $L_1$ off
the right edge of the plot, at very large values of $L/L_*$.

If $L_1<L_{\min}$, the cutoff of the luminosity function,
multiple images cannot be produced. To find out when this situation
occurs, we reexpress the first of equations~(7) as
$$x_c={r_cc^2/2\pi v^2R_0\over D_LD_{LS}/D_SR_0}\,,\eqno{(13)}$$

\noindent where $R_0\equiv c/H_0$ is the Hubble radius. We have isolated
in the numerator and denominator the dependences upon the galactic
model and the cosmological model, respectively (the dependences of
$r_c$ and $R_0$ upon $H_0$ cancel out in the numerator). The denominator of
equation~(13) is plotted, versus the redshift $z_L$ of the lensing
galaxy, in Figure~7 (solid curves) for the various cosmological models.
Multiple imaging can only occur if $x_c<1$, or equivalently, if the
numerator in equation~(13) is smaller than the denominator.
Since $\gamma<1/2$, $r_c/v^2$ is an increasing function of $L$,
and takes its smallest value for $L=L_{\min}$, which was set
to $0.01L_*$ in our simulations. The dotted lines in Figure~7 shows the
resulting minimum values of the numerator of equation~(13) for the various
morphological types. Multiple imaging can only occur when the solid
curve is above the dotted line, in which cases $L_1>L_{\min}$, and $x_c<1$
for all galaxies in the interval $[L_{\min},L_1[$.
As we see, early-type galaxies located at
almost any redshift can produce multiple
images as long as their luminosities are low enough. Spiral galaxies located
too close to the observer, $z_L\simless0.05$, or too close to the
source, $z_L\simgreat2.0-2.4$, cannot produce multiple images.\footnote{$^8$}
{This at
result depends on the particular choice we made for the luminosity
cutoff $L_{\min}$, but notice that 
$(r_c/v^2)_{\min}\propto L_{\min}^{1-2\gamma}$ is a 
weakly-varying function of $L_{\min}$.} Of course, this is a consequence
of our assumed galactic model, which is only an approximation. We do not expect
spiral galaxies to follow rigorously equation~(3), and spiral galaxies at
low redshift can produce multiple images if their core radii happen to
significantly smaller than our model assumes. A spectacular example is
the quadruple-image lens Q2237, which is caused by a lensing spiral
galaxy located at redshift $z_L=0.04$.

\epsfxsize=11cm
\vskip-0.9cm
\hskip2cm\epsfbox{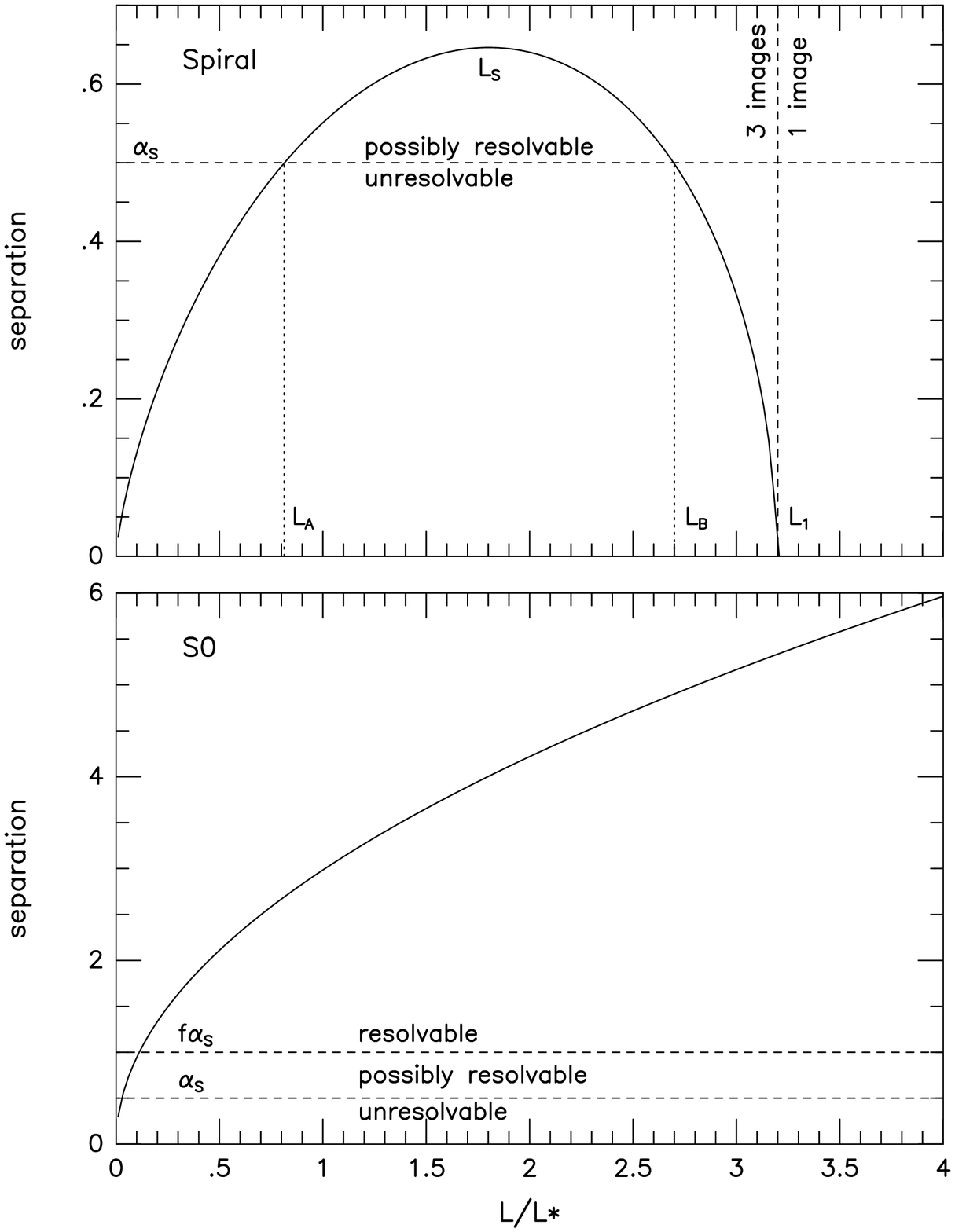}
\vskip-0.8cm

{\narrower\noindent
Fig. 6: Image separation versus galaxy luminosity, for a lensing
spiral galaxy (top panel) and a S0 galaxy (bottom panel) located
at redshift $z=0.3$, in a universe with $\Omega_0=0.3$,
$\lambda_0=0.7$, $H_0=70\rm\,km\,s^{-1}Mpc^{-1}$ (solid curves). 
The horizontal dashed
lines indicate the transitions between the various ``resolvability''
cases described by equation~(5). The vertical dash line, defined by
$x_c=1$, indicates the transition from 3 images to 1. Lensing produces
3 images if $L<L_1$, and these images can be resolved individually
if $L_A<L<L_B$. $L_s$ is the luminosity for which the separation is maximum.
In the bottom panel, $L_A=0.03$ is near the origin;
$L_s$, $L_B$, and $L_1$ are off the right edge of the plot.
\bigskip\smallskip}

Even if a galaxy can produce multiple images, these images might blend
together and be unresolvable if the image separation is too small.
Our model assumes that
images with angular separations less than $\alpha_S=0.5''$ cannot be
resolved individually. The luminosity $L_s$ for which the image separation
is maximum is a very complicated function, but with some simplifying
approximations we can obtain a crude 
estimate of the maximum
image separation that each type of galaxies can produce,
as follows. Using equations~(6) and (7), we rewrite
equation~(10) as
$$s={2\over D_L}\left[4\pi^2\left({v\over c}\right)^4{D_L^2D_{LS}^2\over D_S^2}
-r_c^2\right]^{1/2}\,.\eqno{(14)}$$

\epsfxsize=16cm
\vskip-2.2cm
\hskip0cm\epsfbox{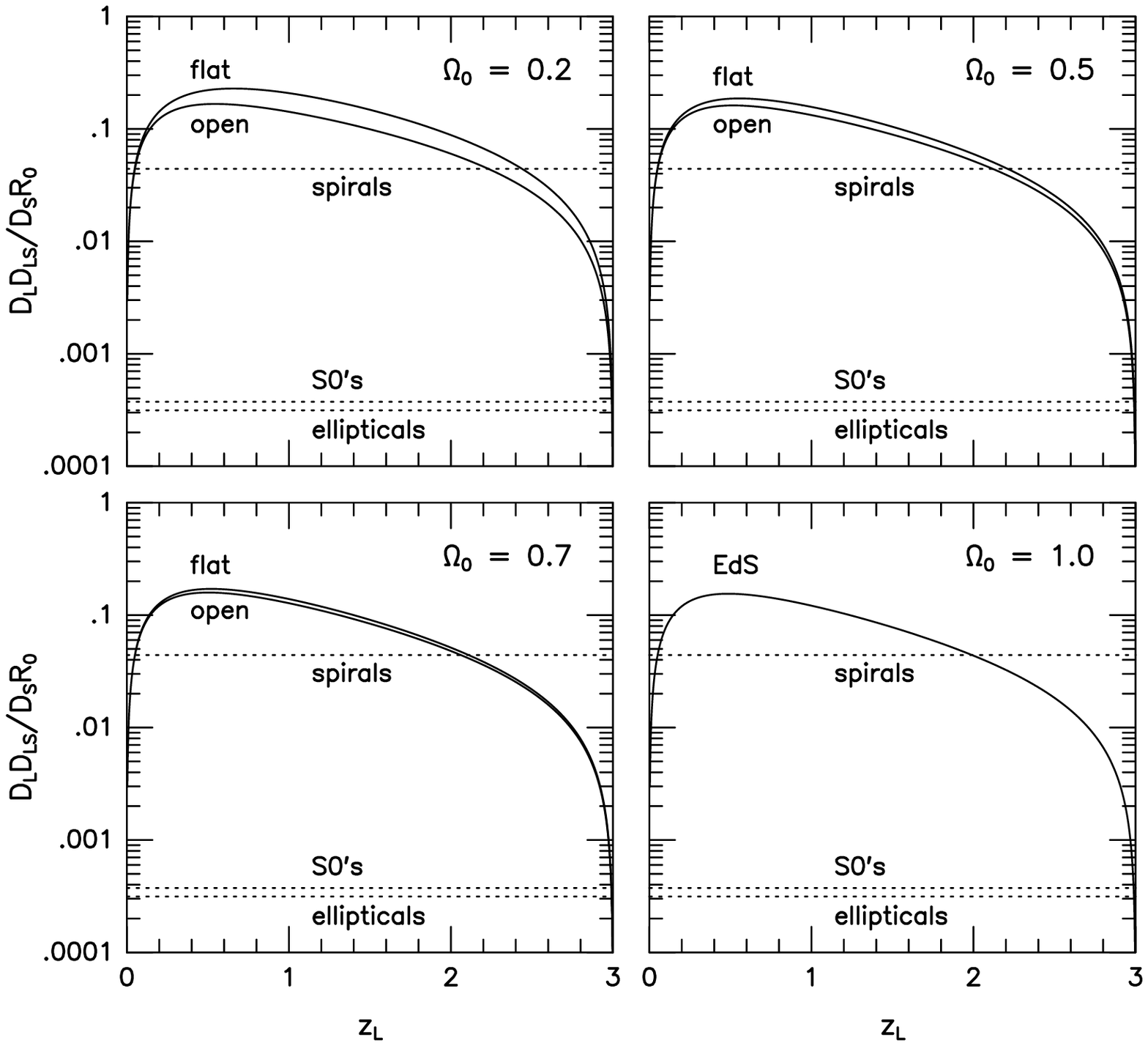}
\vskip-6.8cm

{\narrower\noindent
Fig. 7: Dimensionless ratio $D_LD_{LS}/D_SR_0$ versus lens
redshift $z_L$, for source located at $z_s=3$ (solid curves).
The values of $\Omega_0$ are indicated in each panels. The
curves labeled ``flat'' correspond to $\Omega_0+\lambda_0=1$ models and
the ones labeled ``open'' correspond to $\lambda_0=0$ models. The dotted
lines show the minimum value of the numerator of equation~(13), for
each morphological type.
\bigskip\smallskip}

\noindent Over most redshifts $z_L$ of interest (that is, not too close to
either the observer or the source), the quantities $D_L$ and $D_LD_{LS}/D_S$
are not very sensitive to the redshift or the cosmological parameters
(see Fig.~7 of this paper and Fig.~5 of Paper~II).
In this crude treatment, we can approximate them as constant.
We set $D_L\sim0.3R_0$ and $D_LD_{LS}/D_S\sim0.2R_0$. We also
eliminate $r_c$ and $v$ using equations~(3) and ~(4). Equation~(14)
reduces to
$$s={20\over3}\left[0.16\pi^2\left({v_0\over c}\right)^4
\left({L\over L_*}\right)^{4\gamma}
-{r_{c0}^2\over R_0^2}\left({L\over L_*}\right)^2\right]^{1/2}\,.\eqno{(15)}$$

\noindent This equation depends only upon the galactic model. To find the
maximum separation, we differentiate equation~(15)
relative to $L$, set $ds/dL=0$ and solve for $L$. We get

$${L\over L_*}=\biggl({r_{c0}^2\over0.32\pi^2R_0^2\gamma}\biggr)^{1/(4\gamma-2)}
\biggl({v_0\over c}\biggr)^{-2/(2\gamma-1)}\,.\eqno{(16)}$$

\noindent
We eliminate $L/L_*$ in equation~(15) 
using equation~(16). After some algebra, we get
$$s_{\max}={20\over3}\left({r_{c0}\over R_0}\right)^{2\gamma/(2\gamma-1)}
\left({c^4\over0.32\pi^2\gamma v_0^4}\right)^{1/(4\gamma-2)}
\left({1\over2\gamma}-1\right)^{1/2}\,.\eqno{(17)}$$

\noindent This equation predicts $s_{\max}=93.2''$, $65.5''$, and
$0.8''$ for elliptical, S0, and spiral galaxies, respectively.
Of course, arc-minute separations
like $93.2''$ or $65.5''$ are totally unrealistic. Just like we use a minimum
luminosity $L_{\min}$ in the luminosity function, there should be a maximum
luminosity $L_{\max}$ above which the luminosity function is no
longer valid. Equation~(16) gives $L/L_*=2000$, 1400, and 3.2 for
elliptical, S0, and spiral galaxies, respectively. The first two
numbers are of course absurd. If we choose, say, $L_{\max}=3L_*$ for
all morphological types and substitute this value in equation~(15),
we still get $s_{\max}=5.1''$, $4.2''$, and
$0.8''$ for elliptical, S0, and spiral galaxies, respectively.
Clearly, early-type galaxies totally dominate
the distribution of image separations. 
Spiral galaxies, with their smaller masses and 
much larger core radii, can at best
produce image separations that are merely 60\% larger than the
threshold of resolvability ($s=0.5''$). 

Going back to Figure~6, we indicated on each panel the regions where
individual images are resolvable ($P[s]=1$), unresolvable ($P[s]=0$),
or possibly resolvable ($0<P[s]<1$), according to equation~(5).
For the spiral galaxy (top panel), we identify the following regimes:
$L_{\rm min}<L<L_A$: 3 unresolvable images; $L_A<L<L_B$: 3 resolvable
images; $L_B<L<L_1$: 3 unresolvable images; $L>L_1$: 1 image. Here
$L_A$ and $L_B$ are the values of $L$ for which $s=\alpha_S$. For the
S0 galaxy, the value of $L_A$ is near the origin, at $L_A\sim0.03L_*$,
and the value of $L_B$ is off the right edge of the plot.

These results indicate that, in our model, most spiral galaxies 
responsible for
multiple imaging have {\it intermediate} luminosities (or masses). 
High-luminosity spiral galaxies have large core radii that 
prevent them from
producing multiple images, while low-luminosity 
spiral galaxies produce images with
small separations, which blend together and are not
resolvable individually. For early-type galaxies, the effect 
of the core radius enters only for values of $L_*$ that are 
unrealistically large. Hence, early-type galaxies can always produce
multiple images, unless they are too close to the source or the observer.
Notice that the cross section for multiple imaging still favors
lenses at intermediate redshifts.

In this discussion, we have expressed the image separation in terms
of the luminosity $L$ of the lensing galaxy, and not its mass $M$. 
Of course, for real galaxies, $L$ is a monotonically increasing function of 
$M$, and therefore all the above statements could be rephrased in terms
of the mass. However, our galactic models are defined in terms of
the luminosity (eq.~[1]--[4]). 
The reason is that the mass of an isothermal sphere
(singular or nonsingular) diverges, unless a cutoff radius $r_{\max}$
is introduced in the model. We actually used a cutoff radius $r_{\max}$ in
Papers~I and~II, but for the analytical model presented in this paper,
a cutoff radius is unnecessary. Because the galactic models have a
circularly symmetric projected surface density on the lens plane, 
lensing depends only on the
total mass inside a cylinder of radius equal to the impact parameter on
the lens plane. In the limit $r_{\max}\gg r_c$, this mass is finite and
depends only on $r_c(L)$, $v(L)$, and the impact parameter. Hence, as
long as the edge of each galaxy on the lens plane lies outside its
critical curve, the value of $r_{\max}$, and consequently the value of
the mass $M$, are irrelevant.

\bigskip\smallskip

\ctr{\bf 5.\quad RESULTS}

\medskip

\ctr{5.1.\quad Multiple-Imaging Probability}

\medskip

Using equation~(12), we computed the multiple-imaging probability
$P_{\rm m.i.}$ for the 7 models considered in this paper. The results
are listed in the second column of Table~4. For comparison, we list
in the third column of Table~4 the values obtained in Paper~II, using
ray-tracing experiments. The results are comparable for all models. The
largest difference (19\%) occurs for the Einstein-de~Sitter model E,
but this is precisely the model for which the ray-tracing value is 
most uncertain. These numbers validate our particular
choice of $\zeta=0.5$ in equation~(9). We experimented with the value
of $\zeta$, and found that $\zeta=0$ produced 
analytical values of $P_{\rm m.i.}$ that
were too small by a factor of $2-2.5$ compared with the
ray-tracing values, while $\zeta=1$ produced values that were too large by
a factor of 1.5.

\bigskip\smallskip

\ctr{Table 4: Multiple-Imaging Probability}

\ctr{\vbox{\halign{
\strut#\hfil&\quad\hfil#\hfil&\quad\hfil#\hfil\cr
\noalign{\bigskip\hrule\smallskip\hrule\medskip}
Model & Analytical & Ray-tracing \cr
\noalign{\smallskip\hrule\smallskip}
O1 & 0.0023 & 0.0027 \cr
L1 & 0.0090 & 0.0086 \cr
O2 & 0.0026 & 0.0028 \cr
L2 & 0.0044 & 0.0043 \cr
O3 & 0.0025 & 0.0022 \cr
L3 & 0.0033 & 0.0031 \cr
E  & 0.0023 & 0.0019 \cr
\noalign{\smallskip\hrule}
}}}

\bigskip

The analytical results listed in the second column of Table 4 are
exact. The numerical results listed in the third column are based
on ray-tracing experiments, and are therefore approximate. It is
difficult to estimate the uncertainty on these values, and a detailed
error analysis of the ray-tracing experiments is beyond the scope of this
paper (for details on how these numbers were
obtained, we refer the reader to Paper~II). 
We estimate that the uncertainties range
from 15\% to 25\%, which is large enough to
reconciliate the experiments with the analytical model. 

\bigskip\smallskip

\ctr{5.2.\quad Distributions of Image Separations}

\medskip

\ctr{\sl 5.2.1.\quad Analytical Model vs. Ray-Tracing Experiments}

\medskip

In Figure~8, we compare the the distributions of image
separations predicted by the analytical model (dotted lines) to
the ones obtained in Paper~II using ray-tracing experiments (solid
lines), for the models O1, L1, and E. The agreement is excellent for 
model E (bottom panel). The large peaks at $s=1.9''$ and $s=2.3''$
in the ray-tracing distributions are statistical 
fluctuations. The agreement is quite good for model L1 (middle panel). The
model overestimates the number of cases with $s<1.2''$ and
underestimates the number of cases in the range $1.2''<s<2.3''$ relative to
the ray-tracing experiments. These experiments
take the effect of the background matter
into account, and this tends to shift separations to higher values.
The agreement is not particularly good for model O1 (top panel),
where the ray-tracing experiments
produce a large deficit in the range $0.5''<s<1.2''$, and a large
excess in the range $2.2''<s<3.2''$ relative to the analytical
model. But between these features, the 
analytical model reproduces the ray-tracing results fairly well.

A strange feature that we reported in Paper~II was the presence of a
secondary peak at large separations. We clearly see this peak at
$s\sim5.2''$ in the middle panel, and we could also argue that 
the excesses predicted by the ray-tracing experiments at $s=3.6''$
in the bottom panel and $2.2''<s<3.2''$ in the top panel are manifestations
of the same phenomenon. As we showed in \S3, this secondary peak appears
to be caused by the background matter, in a way that remains to be
explained. The poor agreement between the analytical
model and the simulations for model O1 (top panel), would then be caused
by the particularly large height of this secondary peak.

It is worth pointing out that for a given source size and source redshift,
our analytical model has only two
free parameters: $f$ (eq.~[5]) and $\zeta$ (eq.~[9]). With the particular
combination $f=2$ and $\zeta=0.5$, the model reproduces the distributions
of image separations for 3 very different cosmological models, and
also reproduces the multiple-imaging probabilities shown in Table~4 for
7 different cosmological models. That a 2-parameter model can satisfy so
many constraints supports the notion that this model is accurate and
based on assumptions that are sound.\footnote{$^9$}{Actually, it only shows
that assumptions~(1), (2), (4), and (5) in \S4.2 are sound. 
The analytical model and the ray-tracing simulations both assume the same
galactic models, and therefore assumption (3),
that these galactic models are correct, is neither supported nor 
invalidated by the comparison.}

\epsfxsize=12cm
\vskip-0.6cm
\hskip1.2cm\epsfbox{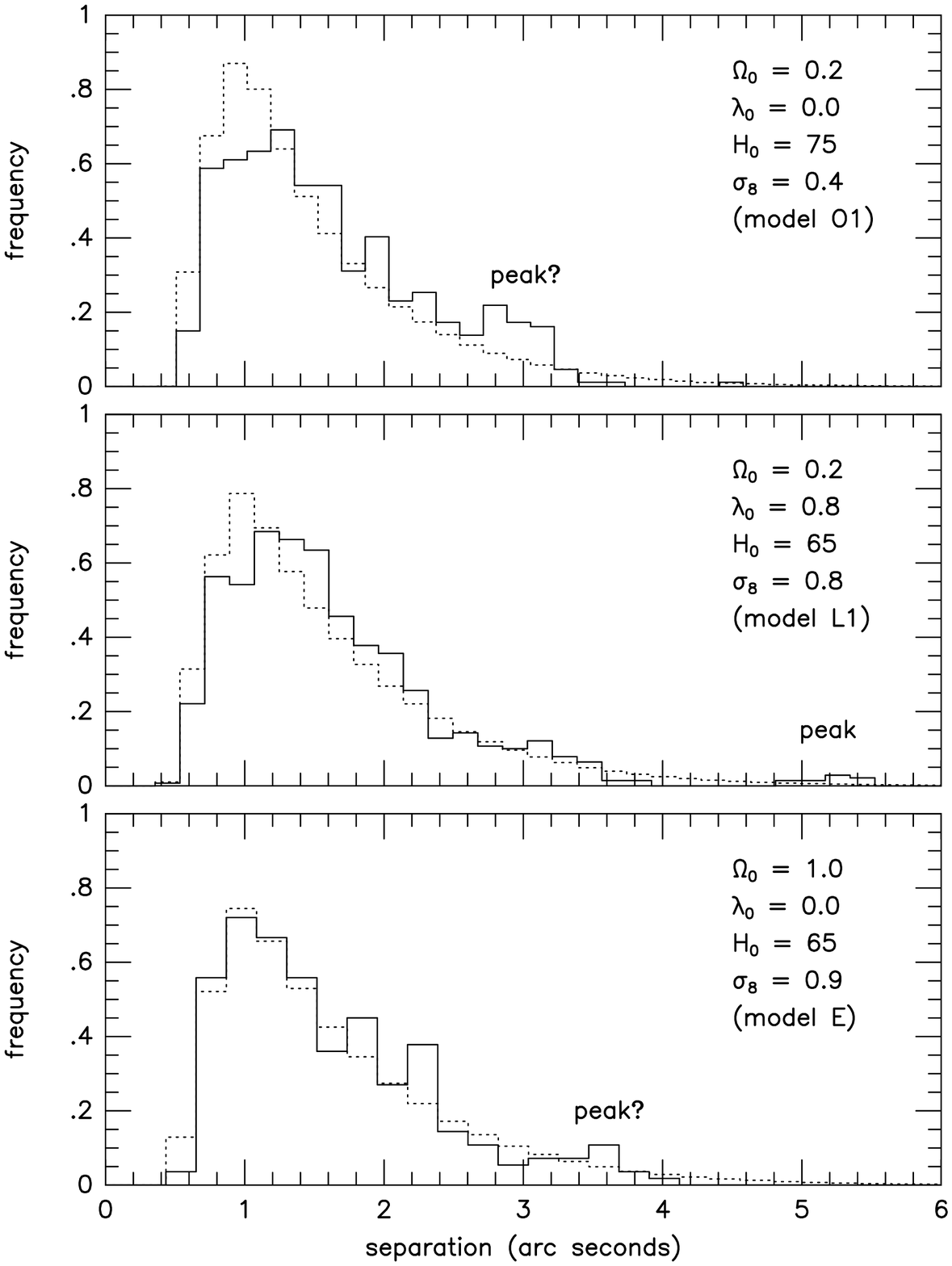}
\vskip-0.6cm

{\narrower\noindent
Fig. 8: 
Distribution of image separations in arc seconds,
computed using ray-tracing experiments (solid lines)
and using the analytical model presented in this paper (dotted lines).
{\it Top}, model O1 
($\Omega_0=0.2$, $\lambda_0=0.0$, $H_0=75\rm\,km\,s^{-1}Mpc^{-1}$,
$\sigma_8=0.4$);
{\it middle}, model L1 
($\Omega_0=0.2$, $\lambda_0=0.8$, $H_0=65\rm\,km\,s^{-1}Mpc^{-1}$,
$\sigma_8=0.8$);
{\it bottom}, model E 
($\Omega_0=1.0$, $\lambda_0=0.0$, $H_0=65\rm\,km\,s^{-1}Mpc^{-1}$,
$\sigma_8=0.9$). The locations or possible locations of
secondary peaks are indicated.}

\bigskip\smallskip

\ctr{\sl 5.2.2.\quad Dependence upon the Cosmological Model}

\medskip

We plot the distributions of the scaled core radii $x_c$, critical
radii $y_r$, and image separations $s$ in Figures~9 and~10. Since
our galaxy distributions, taken from Paper~II, contained over $100,000$
galaxies per model, with luminosities distributed according to equation~(1),
these curves are essentially {\it exact} predictions of the analytical model.
In Figure~9, we compare models E (solid curves) and O1 (dotted curves),
which are both matter-dominated ($\lambda_0=0$), while in Figure~10 we
compare models E (solid curves) and L1 (dotted curves), which are both flat
($\Omega_0+\lambda_0=1$). 

The top panels of Figures~9 and~10 show bimodal distributions, with most
values of $x_c$ being in the range $\sim[0,0.04]$ for early-type galaxies and
$\sim[0.3,1]$ for spirals galaxies. We excluded values of $x_c$ larger than
unity, since they do not result in multiple images. The concentration of
the values of $x_c$ near zero for early-type galaxies implies that, in
our model, these galaxies behave nearly as {\it singular} isothermal
spheres, with the presence of a finite-density core having little
effect. For spiral galaxies, the effect of the finite core can be
very important, and even prevent the formation of multiple images
(when $x_c>1$). Early-type galaxies produce larger image separations than
spirals: not only the factor $\xi_0$ in equation (10) is about 4 times
larger for early-types than spirals (at a given luminosity), 
but in addition the factor
$(1-x_c^2)^{1/2}$, which is always near unity for early-type galaxies, can
be quite lower for spirals. The observational selection effect described
by equation~(5) then becomes very important. As Table~1 shows, our galaxy
distributions contain similar numbers of early-types and spiral galaxies,
and the requirement $x_c<1$ for multiple imaging only eliminate a small
fraction of the spirals. However, the condition given by equation~(5)
eliminates about 50\% of the early-type galaxies, but 95\% of the
spiral galaxies. In other words, only one of every 20 spiral galaxies
can produce images that is sufficiently separated to be
resolved individually. Hence, we expect early-type galaxies
to dominate the distribution of image separations, with a negligible 
contribution from spiral galaxies.

\epsfxsize=14cm
\vskip-1.2cm
\hskip0.4cm\epsfbox{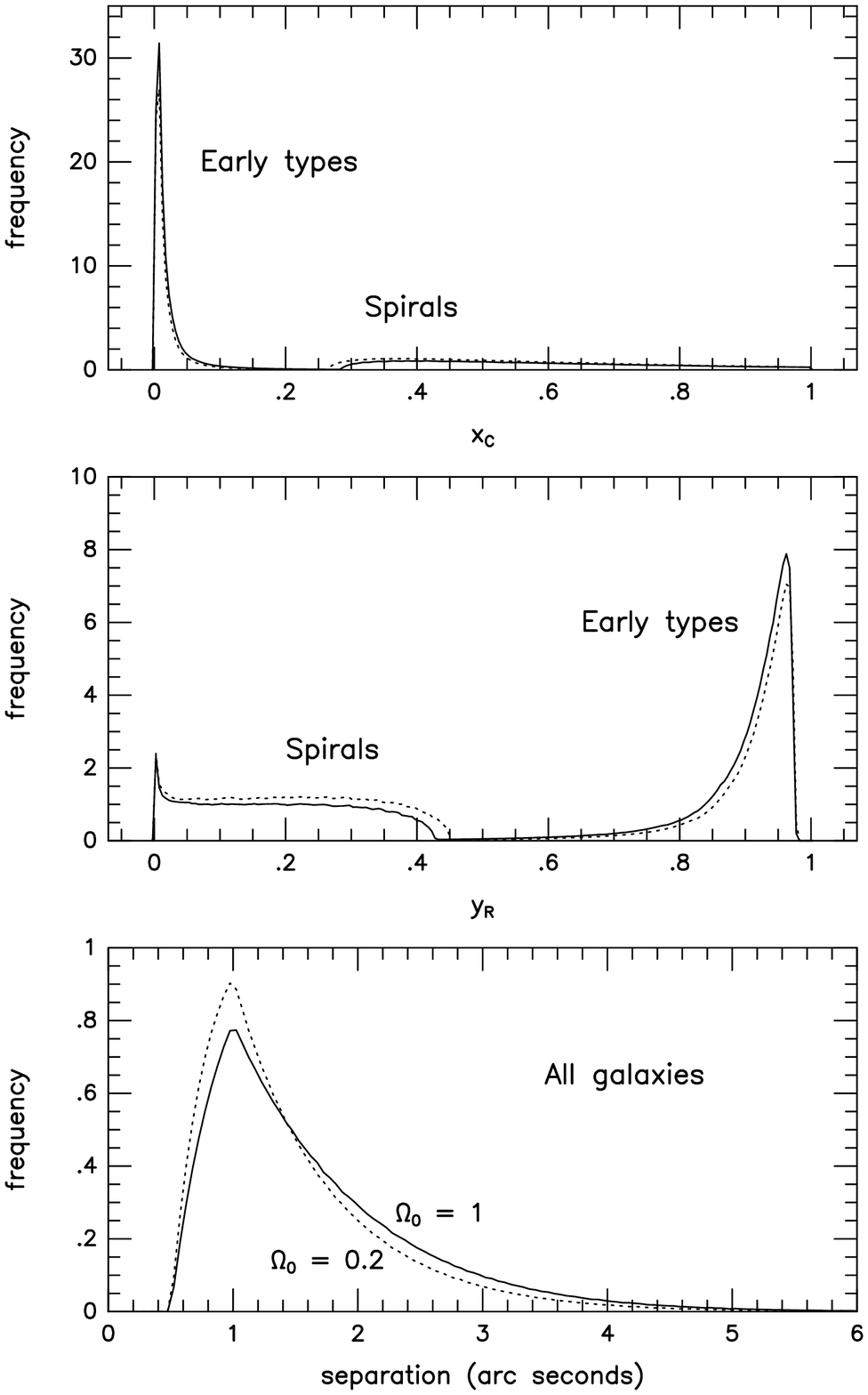}
\vskip-1.2cm

{\narrower\noindent
Fig. 9: {\it Top}, distribution of values of $x_c=r_c/\xi_0$; {\it middle},
distribution of values of $y_r=(1-x_c^2)^{1/2}$; {\it bottom}, distribution of
image separations in arc seconds. The solid and dotted curves correspond to
the $\Omega_0=1$, $\lambda_0=0$ model and $\Omega_0=0.2$, $\lambda_0=0$
model, respectively. Only galaxies capable of producing multiple
images ($x_c<1$) are included.
\bigskip\smallskip}

The second panels of Figures~9 and~10 show the distributions of the values of
$y_r$. The bimodality of the $x_c$ distributions results in the distributions
of $y_r$ being also bimodal, with the values being concentrated near $y_r=1$
for early-type galaxies and spread between 0 and 0.4 for spiral galaxies.
Since the cross section for multiple-imaging scales like $y_r^2$, this
reinforces even more the dominance of early-type galaxies over spirals. 

\epsfxsize=14cm
\vskip-1.2cm
\hskip0.4cm\epsfbox{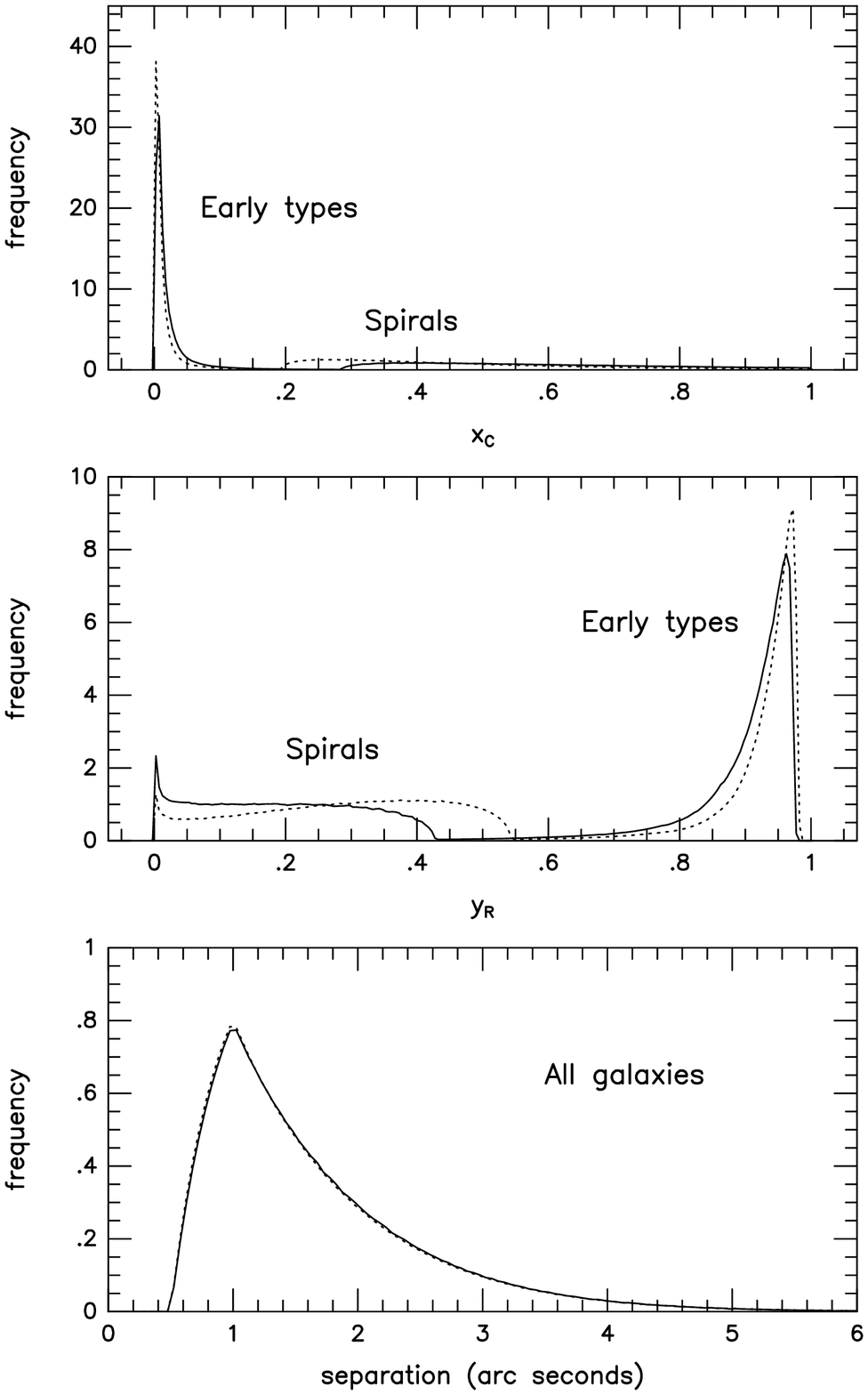}
\vskip-1.2cm

{\narrower\noindent
Fig. 10: Same as Figure~9, except that the dotted curves now correspond
to the $\Omega_0=0.2$, $\lambda_0=0.8$ model. The two curves in the bottom
panel are essentially undistinguishable.
\bigskip\smallskip}

The bottom panels of Figures~9 and~10 show the distributions of
image separations predicted by our analytical model. These distributions
were obtained by computing the image separation for all galaxies using 
equation~(10), and ascribing to each separation a statistical weight $w$
given by equation~(11).
Figure~9 shows that for matter-dominated models ($\lambda_0=0$), 
the distribution depends upon the density parameter. As $\Omega_0$
increases, the peak of the distribution is lowered, and the high-separation
tail extends to larger values. The intermediate 
models O2 and O3 follow this trend. For
clarity, we did not included them in Figure~9.
Figure~10 shows that for flat models ($\Omega_0+\lambda_0=1$), 
the distribution does not depend
upon the density parameter. Indeed the distributions for the E and L1 models,
and for the intermediate models L2 and L3 (not plotted) are nearly
indistinguishable, {\it even though the distributions of values of
$x_c$ and $y_r$ are clearly different}, as the top and middle
panels of Figure~10 show. These differences are actually much more pronounced
than the corresponding ones for the matter-dominated models (top and
middle panels of Fig.~9).

To explain these surprising results, consider first the central panels of
Figures~9 and~10, which show the distributions of the values of $y_r$.
Early-type galaxies have significantly larger values of $y_r$ than
spirals. Since the cross section $\sigma_{\rm m.i.}$ is proportional to
$y_r^2$, early-type galaxies dominate, and we can neglect, to a very good 
approximation, the presence of spiral galaxies. The top panels of
Figures~9 and~10 then show that the values of $x_c$ for early-type
galaxies are much smaller than unity. 
Making the approximation $x_c\approx0$, $y_r=(1-x_c^{2/3})^{3/2}\approx1$,
the expressions for the image separation and the multiple-imaging 
cross section, equations~(10) and~(11), reduce to
$$\eqalignno{
s&\approx4\pi\left({v\over c}\right)^2{D_{LS}\over D_S}\,,&(18)\cr
\sigma_{\rm m.i.}&\approx4\pi^3\left({v\over c}\right)^4
\left({D_LD_{LS}\over D_S^2}\right)^2\,,&(19)\cr}$$

\epsfxsize=14cm
\vskip-2cm
\hskip0cm\epsfbox{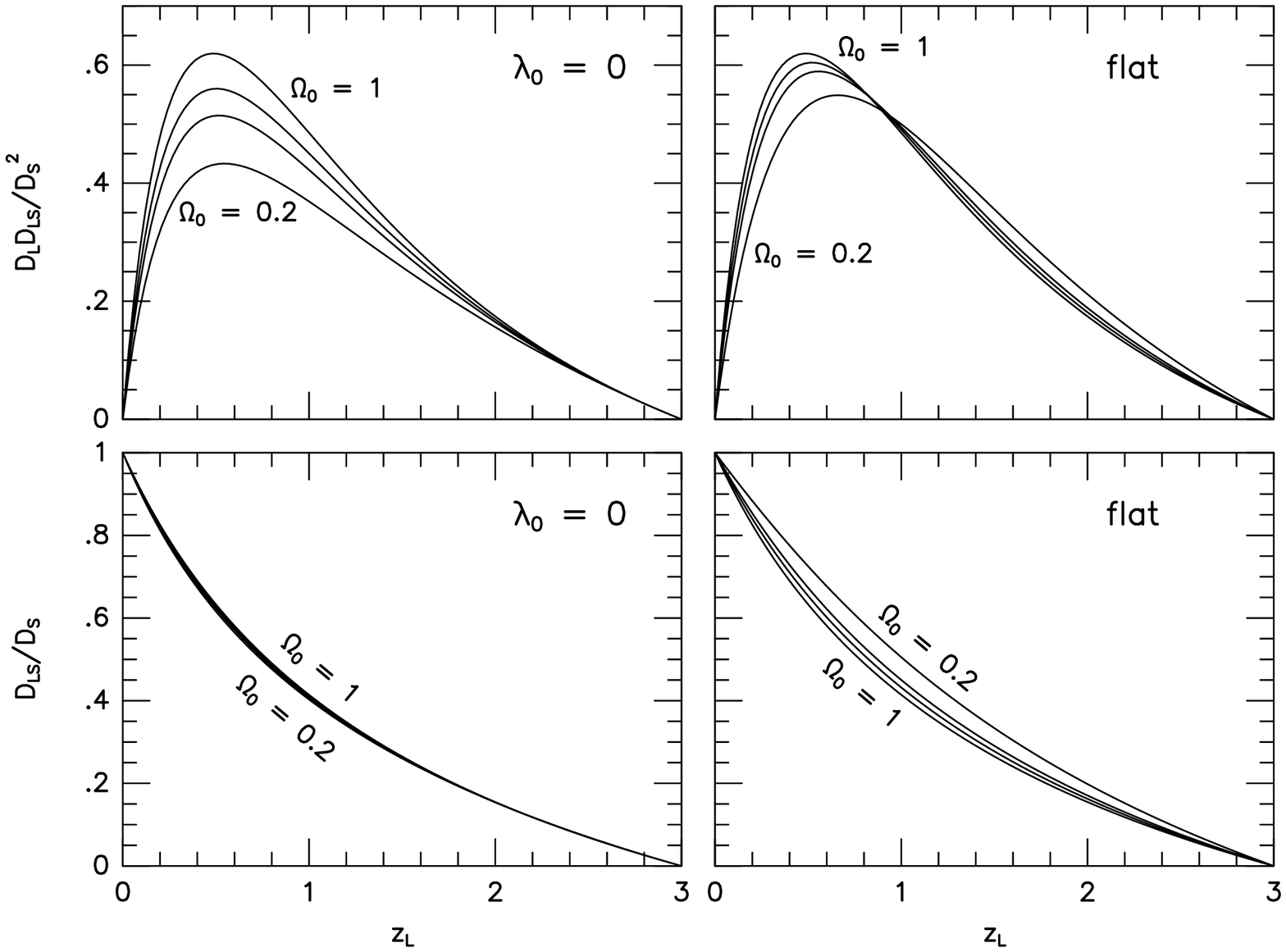}
\vskip-6cm

{\narrower\noindent
Fig. 11: 
Dimensionless angular diameter distance ratios $D_LD_{LS}/D_S^2$
(top panels) and $D_{LS}/D_S$ (bottom panels) versus lens redshift $z_L$,
for sources located at $z_S=3$. Left panels: 
matter-dominated models ($\lambda_0=0$).
Right panels: flat models ($\Omega_0+\lambda_0=1$).
The four curves on each panel correspond to $\Omega_0=0.2$, 0.5, 0.7, and 1.
\bigskip\smallskip}

\noindent
where, in equation~(19), we also neglected the effect of the source size.
These expressions depend upon the cosmological parameters only through the
dimensionless distance ratios $D_{LS}/D_S$ and $D_LD_{LS}/D_S^2$.
These two ratios are plotted versus the lens redshift $z_L$ in
Figure~11. 
Using this figure, we can attempt to interpret the results shown in
Figures~9 and~10. Consider first the matter dominated models
(that is, Fig.~9 and left panels in Fig.~11). The ratio
$D_{LS}/D_S$ is essentially independent of $\Omega_0$. The ratio 
$D_LD_{LS}/D_S^2$, which determines the cross section,
decreases with decreasing $\Omega_0$. The
lensing probability is then reduced for smaller $\Omega_0$, but this effect 
alone could not affect the distributions shown in Figure~9, which are
normalized. Notice, however, that the sensitivity of $D_LD_{LS}/D_S^2$
upon variations of $\Omega_0$ depends on the galaxy redshift $z_L$. Near
$z_L=0.5$, where $D_LD_{LS}/D_S^2$ is maximum, $D_LD_{LS}/D_S^2$ decreases
quite rapidly with decreasing $\Omega_0$, while at redshifts $z_L>1.5$ the
effect is much weaker. Looking now at the bottom left panel of Figure~11,
we see that galaxies located at redshift $z_L\sim0.5$ tend to
produce larger image separations than galaxies located at redshifts
$z_L>1.5$.\footnote{$^{10}$}{But we must keep in mind that the separation does
not depend only on the ratio $D_{LS}/D_S$, but also on the
ratio $v/c$, which has a wide distribution.} Hence, the reduction in cross
section $\sigma_{\rm m.i.}$ resulting from a reduction of $\Omega_0$ affects
large separations more that small separations, explaining the 
effect we see in the bottom panel of Figure~9.

The situation for the flat models (Fig.~10 and right panels in Fig.~11)
is significantly more complicated. First, there is an inversion of
the relationship between $D_LD_{LS}/D_S^2$ and $\Omega_0$. At redshifts
$z_L>0.9$, $D_LD_{LS}/D_S^2$ actually increases with decreasing
$\Omega_0$, and this should reinforce the effect of favoring small
separation angles over large ones. However, as $\Omega_0$ decreases, the 
ratio $D_{LS}/D_S$ increases, resulting in larger separations. These two
effects clearly act in opposite directions. There is one additional 
complication: since all image separations increase with decreasing
$\Omega_0$, several cases with small image separations will be pushed 
above the threshold of resolvability $\alpha_S$
(see eq.~[5]). As Figure~10 shows, these various
effect conspire to produce distributions of image separations 
that are essentially independent of $\Omega_0$.

\epsfxsize=14cm
\vskip-5cm
\hskip1cm\epsfbox{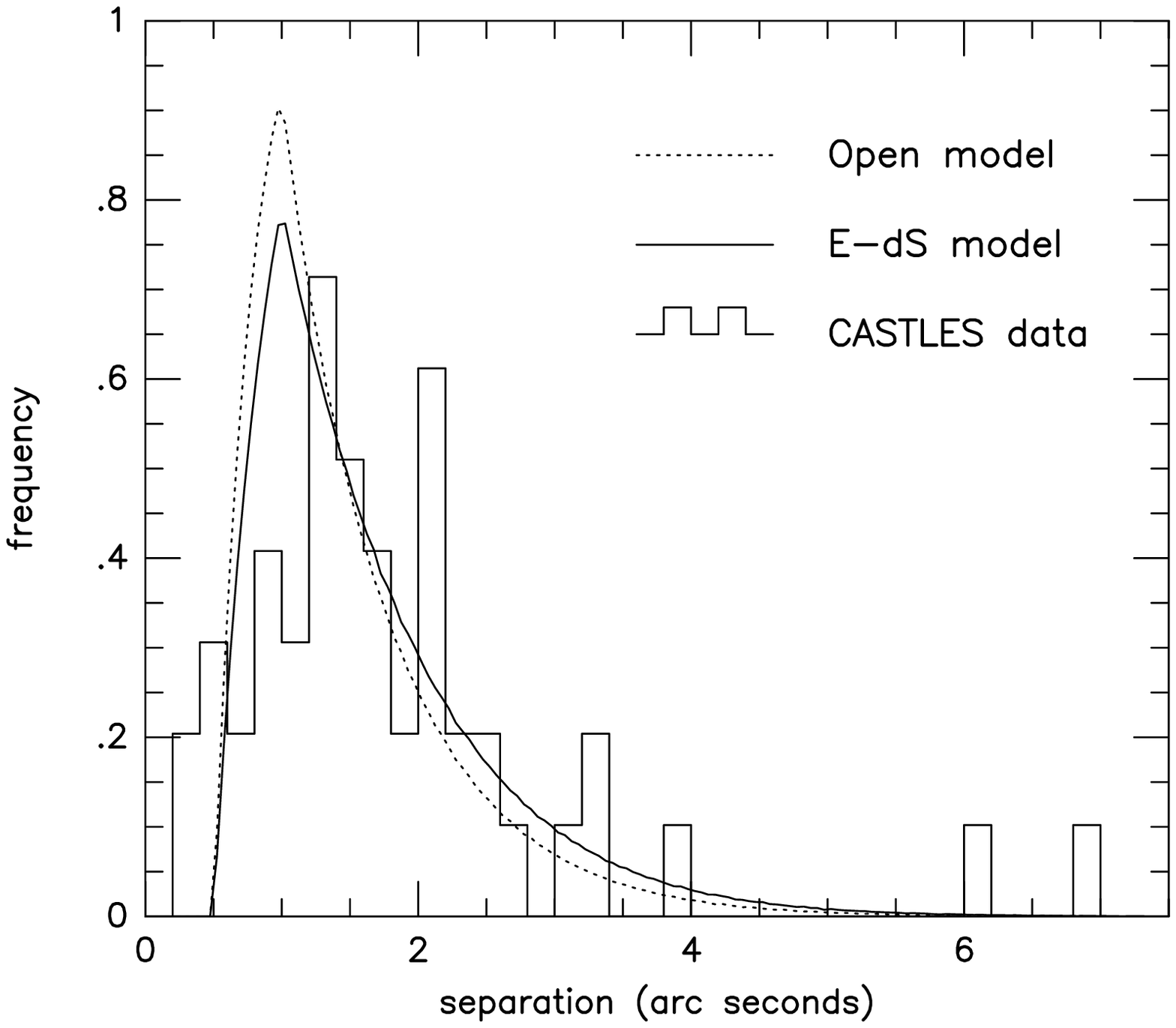}
\vskip-3.5cm

{\narrower\noindent
Fig. 12: Distributions of image separations for the $\Omega_0=0.2$,
$\lambda_0=0$ model (open model, dotted curve)
and $\Omega_0=1$, $\lambda_0=0$ model (Einstein-de Sitter model,
solid curve). The histogram shows the distribution of observed
gravitational lenses, based on the CASTLES database. The bins have
a width of $0.2''$.}

\bigskip\smallskip

\ctr{\sl 5.2.3.\quad Comparison with Observations}

\medskip

In Figure~12, we reproduce the distributions of image separations plotted
in the bottom panel of Figure~9 (solid and dotted curves).
The histogram shows the observed distribution of image separations for 
49 known gravitational lenses, taken from the CASTLES 
database\footnote{$^{11}$}{The
CASTLES database is a web site which lists all information about
known gravitational lenses, obtained from several published surveys
performed by numerous authors.}
(Kochanek et al. 1998). The overall agreement between the models and 
the observations is quite good. The 3 main differences are (1) the peaks
of the distributions are located at $s=1''$ for the models, and $s=1.3''$
for the observations, (2) there are lenses with separations
$s<0.5''$ in the observations, but not in the models, and
(3) two known lenses have
separations $s>6''$, in conflict with the model, which predicts that
such cases are extremely rare. Points~(1) and~(2) are clearly a consequence
of equation~(5), and the fact that our analytical model
assumes a constant source size. 
The rise of the distribution of image separations
at $s=0.5''$ would certainly be less steep if we used instead
a distribution of source sizes, and also a distribution of source redshifts.
Point~(3) is very interesting. There are
currently no observed lenses with image separations in the
range $4''<s<6''$, but there are two lenses with separation $s>6''$,
Q0957+561 (e.g. Young et al. 1980) and RX J0912+4529 (Mu\~noz et al. 2001).
In both cases it is believed that the large image separations result
from the presence of the cluster in which the lens is located.
Such large separations are 
inconsistent with our analytical model which ignores the presence
of the background matter, but consistent with our ray-tracing
experiments when these experiments take the background matter into account.
Could these lenses with $s>6''$ constitute the secondary peak seen in the
middle panels of Figures~1 and~8? This is an exciting possibility, but 
clearly we need more data in order to test this hypothesis. The main conclusion
we draw from Figure~12 is that the analytical model is in good agreement
with the observations, for all cosmological models considered.
A much larger number of observed lenses would be
required in order to rule out the model, and an even larger number would 
be necessary to distinguish between the various cosmological models.

\bigskip\smallskip

\ctr{\bf 6.\quad DISCUSSION}

\medskip

In this section, we review some of the assumptions of the model, and discuss 
their validity.

\bigskip\smallskip

\ctr{6.1.\quad The Galactic Models}

\medskip

Our galactic models (\S4.1) assume that lenses can be approximated as
nonsingular isothermal spheres with a finite core.
While observations of dwarf galaxies and clusters of galaxies
indicate the presence of a flat, constant-density core, HST 
observations of nearby massive galaxies indicate that these
objects also have a core, but the density profile inside these
cores is still singular (or ``cuspy'').
The power-law exponents
vary from 0 (flat core) to 2.5 (steeper than isothermal) 
with a bimodal distribution showing peaks at 0.8 and 2
(Gebhardt et al. 1996). For 
galaxies located in the lower end (exponents less than 1), our 
models can be regarded as a fairly reasonable approximation, 
in any case much better than the commonly used isothermal 
approximation. For galaxies in the upper end, our model is of
course inappropriate. However, Gebhardt et al. (1996)
show that there is a strong correlation between mass and
density profile. The most massive galaxies, which are likely
to be responsible for lensing, are the ones with the flattest
density cores.

In a recent paper, Rusin \& Ma (2001) argue that 
the absence of a detectable third image in lens systems rules out 
shallow profiles. At first sight, this seems to argue against 
profiles with a central core, though in the 
limit of a small core radius, these profiles reduce to an 
isothermal profile.
The lack of detectable odd images rules out 
large regions of shallow profiles for the inner several 
kiloparsecs of galaxy mass distribution. In our galactic models, 
the core radii are given by $r_c=r_{c0}L/L_*$ [eq. (3)] where $r_{c0}$ 
is $1.0h^{-1}\,\rm kpc$ for spirals and $0.1h^{-1}\,\rm kpc$ for early types, 
as indicated in Table 3. Therefore, only spiral galaxies with luminosities 
significantly larger than $L_*$ have a core radius of several 
kiloparsecs, and such galaxies are exponentially rare. In other words,
Rusin \& Ma (2001) rule out the existence of cores that are 
much larger than the ones we consider.

There are other possible explanations for the absence of the third 
image. First, this image tends to be very close to the optical axis, 
and might be hidden behind the lens itself. Unless the lens is transparent, 
such image could not be seen.\footnote{$^{12}$}{Notice that the images studied
by Rusin \& Ma were obtained at radio wavelengths, and lenses should
be transparent
at these wavelenghts.} Also, the third image 
is usually highly demagnified, and might be too faint to be seen. 
Finally, two images would appear as a single image if their angular 
separation is below the resolution limit.
In Paper II, we performed a 
very large number of ray-tracing experiments, using exactly the 
same galaxy profiles as in this paper. We obtained 10,728 double-image
systems, and only 126 triple-image systems (also, a negligible 
fraction of the Einstein rings we produced had a central spot). 
Hence, the paucity of triple-image systems is not inconsistent with
our assumed galactic model. The simulations presented in Paper II
had a finite numerical resolution, and an image could not be 
detected if it was demagnified by a factor of order 40 or more. We
conclude than in these simulations most third images were indeed 
demagnified by a factor of more than 40. Such large demagnifications
result from the fact that our assumed core radii are small, indeed 
significantly smaller than the ones ruled out by Rusin \& Ma.

The relationships between the galaxy parameters
are complex, and the models described in \S4 are only
approximate. In particular, there is probably no one-to-one
correspondence between core radius, luminosity, and rotation
velocity for real galaxies. Also, our model assumes spherically symmetric
lenses which follow a Schechter luminosity function. In the real
world, lenses are not spherical, and the luminosity function is
more complex than a Schechter form (see, in particular, Keeton
et al. 2000).

In any case, it is very important to keep in mind what the goal 
of this paper is. Image separations can in principle tell us about
either the nature of the lenses or the cosmological model. In this
paper, we consider only the latter. The goal of this paper is not
to make a precise comparison with observations and find out which 
galactic model reproduces observations better. 
The goal is to find out whether image separations can
tell us anything about the cosmological models. The key results of
this paper are shown in the bottom panels of Figures 9 and 10,
and described in \S5.2.2.
The important thing to realize is that these results are comparative. 
The main conclusion of this paper is based on the fact that the 
distributions of image separations are similar in the bottom panel 
of Figure 9, and identical in the bottom panel of Figure 10.
If we had used different galactic models, these curves would
probably have been different. However, the {\it relationship} between
these curves
would most likely be the same: The distributions would still be 
similar in Figure 9, and nearly identical in Figure 10. The reason
is that these relationships are determined essentially by the 
properties of the angular diameter distance ratios and their 
dependence on redshift, as plotted in Figures 7 and 11. Hence,
the particular choice of galactic models is not critical, and
as long as these models are reasonable, the 
conclusions remain valid. 

\bigskip\smallskip

\ctr{6.2.\quad Isolated galaxies}

\medskip

The formulae we use for the angular cross section and image 
separations (eqs. [6]--[10]) were derived for an isolated nonsingular 
isothermal sphere. In this paper, we apply these results to our
galaxies, assuming that each galaxy acts as if it was alone. When
we treat a galaxy as "isolated," we are neglecting both the presence
of nearby galaxies and the presence of background matter, that would
be present if the galaxy is part of a cluster.

The effect of nearby galaxies is likely be small, for 2 reasons:
(1) Even in the center of dense clusters, the number density of
galaxies is sufficiently small that having 2 galaxies along a given line
of sight is unlikely (see Fig. 10 of Paper II), and if galaxies are
off the line of sight, their tidal influence drops rapidly. (2)
The most massive galaxies are the ones usually
responsible for lensing. The galaxies located near one of these
massive galaxies are then presumably much less massive, and their
effect would be at most a small correction.

The background matter is a more tricky issue. TOG
addressed this issue, and derived an estimate of
the effect (their equation [2.36] and [2.37]), but give some words of
caution in their Appendix. If the lensing galaxy is located in the
center of a dense cluster, the presence of the cluster would increase
the image separation. However, the regions located immediately in 
front of or behind the cluster would presumably be underdense, and 
their presence would partly compensate the effect of the cluster.
Still, we were quite concerned about the possible effect of the
background matter. This is why we performed the ``no-background''
ray-tracing experiments described in \S3. These experiments
reveal that the effect of the background matter on the distribution
of magnifications can be very important (see Fig.~2), but the
effect on the distribution of image separations is quite small
(see Fig.~1), a spreading of the high-separation tail toward
higher values, which is similar in all models and therefore does 
not affect the comparison between the models.
This extent of the high-separation tail
of the distribution toward higher separations can explain
observed lenses with separations larger than $6''$ 
(Q0957+561 and J0921+4529).
Since the main conclusion of the paper is based on overall comparisons
between distributions, it is not affected by the statistics of such rare,
very-high-separation cases.

\bigskip\smallskip

\ctr{6.3.\quad Source Size}

\medskip

In our analytical model, we correct the angular cross section to
account for the finite size of the source [eq.~(9)]. Throughout
this paper, we assume a fixed source size of diameter $2\alpha_S=1''$.
We introduced the source 
size in the model after finding a discrepancy between the analytical 
results and the ray-tracing experiments. Since these experiments all 
assumed a source size of 1", we introduced it in the model, and found 
out that the numbers listed in Table 4 were in good agreement when
the parameter $\zeta$ was set to 0.5.
 
Hence, the source size was introduced merely to demonstrate that the
numerical experiments and the analytical model were not in conflict.
But from a theoretical viewpoint, it is interesting to have a model
which takes the source size into account, and one is always free
to set it to zero afterward. Alternatively, the analytical model does not
require sources to have all the same size, and a distribution
of source sizes could be used instead. Changing
the source size (or neglecting it) would modify the various 
distributions of image separations, 
but not their relationships with each others, and
therefore would not affect the conclusion. 

\bigskip\smallskip

\ctr{\bf 7.\quad SUMMARY AND CONCLUSION}

\medskip

We have designed a simple analytical model to study the distribution
of images separations caused by gravitational lensing. Our approach
differs from previous work in that we use a galactic model in
which the radial density profile has a finite core, while most other
studies assume singular density profiles, either singular isothermal
spheres, NFW profiles, or point masses. In addition, our model
considers the finite angular size of the sources. This has
two effects. First, the finite size of the sources introduces an
observational selection effect. Individual images cannot be resolved if
their separation is less than the angular radius of the unlensed source.
Second, the finite size of the sources 
increases the effective cross section for multiple imaging.
A key assumption of our model is that lensing is caused by
galaxies only, and the presence of the background matter can be ignored.
To test the validity of this assumption, we performed a series of
ray-tracing experiments, using the multiple lens-plane algorithm
described in Paper~I. 
Our results are the following:

(1) The presence of the background matter tends to increase the image
separations produced by lensing galaxies. The peak of the distribution
is lowered, the distribution becomes wider, and often develops a
high-separation tail. However, this effect is rather small, of
order 10\%. Furthermore, this effect appears to be independent
of the cosmological model. We considered three very different
cosmological models: an open, low-density model, a flat, cosmological-constant
model, and an Einstein-de~sitter model, and found that the effect of the
background matter on the distribution of image separation is
essentially the same for all models. The effect is always small, and
it is therefore correct to ignore the presence of the background matter
in our analytical model.

(2) Simulations with galaxies and background matter often produce a
secondary peak in the distribution of image separations at large
separations, as we noticed previously in Paper II. 
This peak does not appear in simulations which only include the effect 
of galaxies, and therefore results from some coupling effect between the
galaxies and the background matter.

(3) The effect of the background
matter on the magnification distribution is strongly dependent
on the cosmological model. This effect is completely negligible in low
density models with small density contrast ($\Omega_0=0.2$, $\sigma_8=0.4$),
but becomes very important in models with larger $\Omega_0$ and 
$\sigma_8$ increase, resulting in a significant widening of the distribution.
We speculate that the absence of any effect of the background matter in
the low-density models results from a cancellation of the density 
fluctuations along the line of sight, which does not occur (or only
to a lesser extend) in cosmological models with higher $\Omega_0$. We
will investigate this problem in more detail in a forthcoming paper.

(4) Nearly all multiple images are caused by early-type galaxies
(ellipticals and S0's).
The contribution of spiral galaxies to the distribution of
image separations is totally negligible. The dominance of early-type
galaxies  over 
spirals is an old result that has been pointed out by several authors.
Notice, however, that in all studies that assume a galactic model with a 
singular density profile, the only reason why early-types dominate is because
of their larger 
masses. In our study, galaxies have a finite-density core, and
the difference in core radii between early-types and spirals not only
reinforces the dominance of early-types, but this effect is even more
important than the effect of the mass
difference. Furthermore, imposing an observational selection
effect based on the finite size of the source (eq.~[5]) eliminates about
50\% of the multiple images produced by early type galaxies, but more than
95\% of the ones produced by spirals.

Of course this is only valid within the assumptions of the
analytical model. There are 3 known lenses caused
by spiral galaxies (JVAS B0318+357, CLASS B1600+434, and Q2237+030),
or about 5\% of the total number of observed lenses.
The analytical model assumes a one-to-one
correspondence between core radius and luminosity, while in the
real universe there must be a distribution of core radii at any 
given luminosity. These three lenses must be spiral
galaxies with particularly small core radii.

(5) Without the assumption of a finite source size, our analytical model 
would have no free parameter. The finite size of the source
enters in the model in two different ways. First, it introduces an
observational selection effect, described in terms of a probability $P(s)$
of resolving individually
two images separated by an angle $s$. Second, it increases
the effective cross section for multiple imaging by a galaxy. Each one
of these effects introduces a free parameter in the model: $f$ in
equation~(5), which fixes the maximum separation above which 
two images can always be resolved individually, and $\zeta$ in
equation~(9), which measures the increase in the cross section for
multiple-imaging.\footnote{$^{13}$}{Notice that $\alpha_S$ in eq.~[5], the source 
angular radius, is not a free parameter, but a value that must be chosen 
{\it a priori}, just like the cosmological parameters $\Omega_0$ and 
$\lambda_0$, or the parameters in the galaxy luminosity function.}
By fitting our analytical model to the results of the ray-tracing experiments
presented in Paper~II, we found that with the particular combination
$f=2$, $\zeta=0.5$, our analytical model
successfully reproduces the
distributions of image separations, and also the multiple-image
probability, for all cosmological models considered in this paper. 

(6) The analytical
model predicts that the distributions of image separations are
virtually indistinguishable for flat, cosmological constant models 
($\Omega_0+\lambda_0=1$) with
different values of $\Omega_0$. For models
without a cosmological constant, the distributions of
image separations does depend upon $\Omega_0$, but this dependence
is weak. Using the dependence of the image separation and the
multiple-imaging cross section on the angular diameter distances, we have
come out with a tentative explanation for these results, but this issue
probably needs more investigation.

We have considered 7 very different cosmological models, and found
that these models produce distributions of image separations that are
extremely similar. The largest difference is illustrated in the
bottom panel of Figure~9. 
We conclude that while the number of multiple-imaged sources
can put strong constraints on the cosmological parameters
(see, e.g., Fox \& Pen 2001), the distribution
of image separations does not constrain the cosmological models in any
significant way, and mostly provides constraints on the structure of
the galaxies responsible for lensing.

We have assumed throughout this paper a fixed source redshift $z_S=3$
and a fixed source angular diameter $2\alpha_S=1''$. The source size
determines the location of the peak in the distribution of image
separations. By allowing the
source sizes and/or source redshifts to vary,
we could modify the shape of the distributions of image separations,
especially at small separations.
However, these distributions would certainly remain nearly independent of
the cosmological model, which is the main result of this paper.
Hence, we do not believe that the 
assumptions of fixed source size and source redshift
limit the validity of our results.

\bigskip

We are thankful to Karl Gebhardt for stimulating discussions.
This work was supported by Grant 3658-0624-1999 from the Texas Advanced
Research Program, NASA Grants NAG5-10825 and NAG5-10826,
and Grants NSF ACI 9982297, NSF PHY 0102204, 
and NPACI NSF UCSD 10181410.
PP is very grateful to Rachel Webster and 
the astrophysics group at the University of Melbourne
for the hospitality and fruitful discussion.

\bigskip

%

\parindent=0pt

\def \hh {\hangindent=30pt\hangafter=1}

\ctr{REFERENCES}

\medskip

\hh
Bernstein, G., \& Fischer, P. 1999, AJ, 118, 14

\hh
Blandford, R. D., \& Kochanek, C. S. 1987, ApJ, 321, 658

\hh
Burkert, A., \& Silk, J. 1999, in
Dark Matter in Astrophysics and Particle Physics, Proceedings of the second International Conference on Dark Matter in Astrophysics and Particle 
Physics, eds. H. V. Klapdor-Kleingrothaus and L. Baudis
(Philadelphia: Institute of Physics Publishers), p. 375

\hh
Cheng, Y.-C. N., \& Krauss, L. M. 1999, ApJ, 514, 25

\hh
Cole, S, \& Lacey, C. 1996, MNRAS, 281, 716

\hh
Dressler, A. 1980, ApJ, 236, 351

\hh
Dyer, C. C. 1984, ApJ, 287, 26

\hh
Efstathiou, G., Ellis, R. S., \& Peterson, B. A. 1988, MNRAS, 232, 431

\hh
El-Zant, A., Shlosman, I., \& Hoffman, Y. 2001, ApJ, 560, 636

\hh
Fox, D. C., \& Pen, U.-L. 2001, ApJ, 546, 35

\hh
Fukushige, T., \& Makino, J. 1997, ApJ, 477, L9

\hh
Fukushige, T., \& Makino, J. 2001a, ApJ, 557, 533

\hh
Fukushige, T., \& Makino, J. 2001b, preprint (astro-ph/0108014)

\hh
Gebhardt, K. et al. 1996, AJ, 112, 1

\hh
Ghigna, S., Moore, B., Governato, F., Lake, G., Quinn, T., \& Stadel, J. 2000,
ApJ, 544, 616

\hh
Hinshaw, G., \& Krauss, L. M. 1987, ApJ, 320, 468

\hh
Huss, A., Jain, B., \& Steinmetz, M. 1999, MNRAS, 308, 1011

\hh
Jaroszy\'nski, M. 1991, MNRAS, 249, 430

\hh
Jaroszy\'nski, M. 1992, MNRAS, 255, 655

\hh
Jing, Y., \& Suto, Y. 2000, ApJ, 526, L69

\hh
Keeton, C. R. 2001, ApJ, 561, 46

\hh
Keeton, C. R., Christlein, D., \& Zabludoff, A. I. 2000, ApJ, 545, 129

\hh
Klypin, A., Kravtsov, A. V., Bullock, J. S., \& Primack, J. R. 2000,
preprint (astro-ph/0006343)

\hh
Kochanek, C. S. 1995, ApJ, 453, 545

\hh
Kochanek, C. S., \& Blandford, R. D. 1987, ApJ, 321, 676

\hh
Kochanek, C. S., Falco, E. E., Impey, C.,
Leh\'ar, J., McLeod, B, \& Rix, H.-W. 1998,
CASTLES Gravitational Lens Database (Cambridge: CfA)

\hh
Kravtsov, A. V., Klypin, A. A., Bullock, J. S., \& Primack, J. R. 1998,
ApJ, 502, 48

\hh
Lauer, T. R. et al. 1995, AJ, 110, 6

\hh
Li, L.-X., \& Ostriker, J. P. 2001, preprint (astro-ph/0010432)

\hh
Moore, B., Governato, F., Quinn, T., Stadel, J., \& Lake, G. 1998,
ApJ, 409, L5

\hh
Moore, B., Quinn, T., Governato, F., Stadel, J., \& Lake, G. 1999,
MNRAS, 310, 1147

\hh
Mu\~noz, J. A. et al. 2001, ApJ, 546, 769

\hh
Narayan, R., \& White, S. D. M. 1988, MNRAS, 231, 97p

\hh
Navarro, J. F., Frenk, C. S., \& White, S. D. M. 1996, ApJ, 462, 563

\hh
Navarro, J. F., Frenk, C. S., \& White, S. D. M. 1997, ApJ, 490, 493

\hh
Paczy\'nski, B., \& Wambsganss, J. 1989, ApJ, 337, 581

\hh
Porciani, C., \& Madau, P. 2000, ApJ, 532, 679

\hh
Postman, M., \& Geller, M. J. 1984, ApJ, 281, 95

\hh
Premadi, P., Martel, H., \& Matzner, R. 1998, ApJ, 493, 10 (Paper I)

\hh
Premadi, P., Martel, H., Matzner, R., \& Futamase, T. 2001, 
ApJ Suppl., 135, 7 (Paper II)

\hh
Primack, J. R., Bullock, J. S., Klypin, A. A., \& Kravtsov, A. V. 1999,
in Galaxy Dynamics, eds. by D. R. Merritt, M. Valluri, and J. A. 
Sellwood, ASP Conference Series vol. 182 (San Francisco: ASP)

\hh
Romanowsky, A. J., \& Kochanek, C. S. 1999, ApJ, 516, 18

\hh
Rusin, D. \& Ma, C.-P. 2001, ApJ, 549, L33

\hh
Schneider, P., Ehlers, J., \& Falco, E. E. 1992, Gravitational Lenses
(New York: Springer) (SEF)

\hh
Shapiro, P. R., Iliev, I. T., \& Raga, A. C. 1999, MNRAS, 307, 203

\hh
Spergel, D. N., \& Steinhardt, P. J. 1996, Phys.Rev.Lett., 84, 3760

\hh
Tormen, G. , Bouchet, F. R., \& White, S. D. M. 1997, MNRAS, 286, 865

\hh
Takahashi, R., \& Chiba, T. 2001, preprint (astro-ph/0106176)

\hh
Turner, E. L., Ostriker, J. P., \& Gott, J. R. 1984, ApJ, 284, 1 (TOG)

\hh
Tyson, J. A., Kochanski, G. P., \& Dell'Antonio, I. P. 1998, ApJ, 498, L107

\hh
Walsh, D., Carswell, R. F., \& Weymann, R. J. 1979, Nature, 279, 381

\hh
Young, P., Gunn, J. E., Kristian, J., Oke, B., \& Westphal, J. A. 1980,
ApJ, 241, 507

\vfill\eject\end